\documentclass[twocolumn]{aastex63}
\usepackage{graphicx}
\usepackage{multirow}
\usepackage{amssymb}
\usepackage{endnotes} 
\usepackage{footnote}
\usepackage{amsthm}
\usepackage{amsmath}
\usepackage{amsfonts}
\usepackage{natbib}
\usepackage{url}
\usepackage{booktabs}
\usepackage{inputenc}
\usepackage{isotope}
\usepackage{hyperref}
\usepackage{longtable}
\usepackage{afterpage} 
\usepackage{pifont}
\usepackage{rotating}
\newcommand{\cmark}{\ding{51}}%
\newcommand{\xmark}{\ding{55}}%

\received{}
\revised{}
\accepted{}
\shorttitle{SN~2020sck}
\shortauthors{Dutta et al.}
\graphicspath{{./}{figures/}}

\begin{document}

\title{SN~2020sck: deflagration in a carbon-oxygen white dwarf}

\correspondingauthor{Anirban Dutta}
\email{anirban.dutta@iiap.res.in, anirbaniamdutta@gmail.com}

\author[0000-0002-7708-3831]{Anirban Dutta}
\affiliation{Indian Institute of Astrophysics, II Block, Koramangala, Bangalore 560034, India}

\author{D.K. Sahu}
\affiliation{Indian Institute of Astrophysics, II Block, Koramangala, Bangalore 560034, India}

\author[0000-0003-3533-7183]{G. C. Anupama}
\affiliation{Indian Institute of Astrophysics, II Block, Koramangala, Bangalore 560034, India}

\author{Simran Joharle}
\affiliation{Department of Physics, Fergusson College, Autonomous, Pune}

\author{Brajesh Kumar}
\affiliation{Aryabhatta Research Institute of Observational Sciences, Manora Peak, Nainital 263 001, India}

\author{Nayana A J}
\affiliation{Indian Institute of Astrophysics, II Block, Koramangala, Bangalore 560034, India}

\author[0000-0003-2091-622X]{Avinash Singh}
\affiliation{Hiroshima Astrophysical Science Center, Hiroshima University, Higashi-Hiroshima, Hiroshima 739-8526}

\author{Harsh Kumar}
\affiliation{Department of Physics, Indian Institute of Technology, Bombay, Powai, Mumbai 400076, India}
\affiliation{LSSTC DSFP Fellow 2018}

\author{Varun Bhalerao}
\affiliation{Department of Physics, Indian Institute of Technology, Bombay, Powai, Mumbai 400076, India}

\author{Sudhansu Barway}
\affiliation{Aryabhatta Research Institute of Observational Sciences, Manora Peak, Nainital 263 001, India}





\begin{abstract}

We present optical $UBVRI$ photometry and low-to-medium resolution spectroscopic observations of type Iax SN~2020sck spanning $-$5.5~d to $+$67~d from maximum light in the $B$-band. 
From the photometric analysis we find \(\Delta m_{\rm{B}}(15) = 2.03\pm0.05\) mag and \(M_{\rm{B}}=-17.81\pm0.22\) mag. Radiation diffusion model fit to the quasi-bolometric light curve indicates \(0.13\pm0.02\) \(M_\odot\) of $^{56}$Ni and 0.34 \(M_{\rm \odot}\) of ejecta are synthesized in the explosion. Comparing the observed quasi-bolometric light curve with angle-averaged bolometric light curve of three-dimensional pure deflagration explosion of \(M_{\rm{ch}}\) carbon-oxygen white dwarf, we find agreement with a model in which \(0.16\) \(M_\odot\) of $^{56}$Ni and \(0.37\) \(M_\odot\) of ejecta is formed. By comparing the $+$1.4~day spectrum of SN~2020sck with synthetic spectrum generated using \texttt{SYN++}, we find absorption features due to C\,{\sc ii}, C\,{\sc iii} and O\,{\sc i}. These are unburned materials in the explosion and indicate a C-O white dwarf. One dimensional radiative transfer modeling of the spectra with \texttt{TARDIS} shows higher density in the ejecta near the photosphere and a steep decrease in the outer layers with an ejecta composition dominated mostly by C, O, Si, Fe, and Ni. The star formation rate of the host galaxy computed from the luminosity of the H$\alpha$ ($\lambda$6563) line is 0.09~\(M_{\odot}\)~yr$^{-1}$ indicating a relatively young stellar environment.

\end{abstract}


\section{\textbf{Introduction}} \label{sec:intro}

The thermonuclear supernovae, also known as type Ia supernovae (SNe Ia), result from the thermonuclear explosion of degenerate carbon-oxygen (C-O) white dwarf that accretes matter from a companion star (\citealt{1960ApJ...132..565H, 2012NewAR..56..122W, 2016IJMPD..2530024M}). There exists a tight correlation between luminosity at maximum light and light curve decline rate for normal SNe Ia, with brighter objects having broader light curves \citep{1999AJ....118.1766P}. This points towards the fact that SNe Ia form a homogeneous class of objects. But, observations have revealed that there exists significant diversity that could be due to the progenitor systems and/or the explosion mechanisms (\citealt{2014ARA&A..52..107M, 2017hsn..book..317T}). One such peculiar subclass of thermonuclear origin is the Iax type (SNe Iax). SN~2002cx was the first of this kind studied in detail (\citealt{2003PASP..115..453L, 2006AJ....132..189J}).

The most important distinguishing feature of SNe Iax is their low expansion velocity of 2,000 km~\(\rm{s^{-1}}\) \citep[SN~2008ha,][]{2009AJ....138..376F} - 8,000 km~\(\rm{s^{-1}}\) (SN~2012Z, \citealp{2014A&A...561A.146S}) as compared to SNe Ia \citep[$v$~$\sim$ 11,000 km~\(\rm{s^{-1}}\),][]{2009ApJ...699L.139W, 2013ApJ...767...57F}, measured from the absorption minimum of the P-Cygni profiles. Their pre-maximum spectra are similar to SN~1991T-like objects. The spectra are dominated by ions of C\, {\sc ii}, C\, {\sc iii}, O\, {\sc i}, intermediate mass elements (IME's) like Mg\, {\sc ii}, Si\, {\sc ii},  Si\, {\sc iii}, S\, {\sc ii}, Ca\, {\sc ii}, Sc\, {\sc ii}, Ti\, {\sc ii} and also Fe group elements (IGE's) like Fe\, {\sc ii}, Fe\, {\sc iii}, Co\, {\sc ii}, Co\, {\sc iii}. The late time spectra are dominated with permitted Fe\, {\sc ii} lines with low expansion  velocities (\citealt{2006AJ....132..189J, 2008ApJ...680..580S}). This indicates that the inner regions of the ejecta in SNe Iax have a higher density compared to normal SNe Ia.

SNe Iax tend to be less luminous as compared to SNe Ia ($\sim -$19.3 mag). The peak optical luminosity spans a wide range from \(M_{\rm{V}}\) = -18.4 mag \citep{2011ApJ...731L..11N} to  \(M_{\rm{g}}\) = - 13.8 mag \citep{2020ApJ...892L..24S}. 
The $I$ band light curve does not show the secondary maximum,  typically seen in SNe Ia caused either due to higher ionization of the absorption lines of Fe and Co \citep{2006ApJ...649..939K} or strong mixing in the ejecta, which reduces the Fe-peak elements in the central region \citep{2006A&A...453..229B}. The $(B-V)$, $(V-R)$, and $(V-I)$ color curves show significant scatter, which can be related to host galaxy reddening or intrinsic to the SN itself \citep{2013ApJ...767...57F}.  

Several progenitor systems and their variants have been proposed to understand the nature of the explosion of SNe Ia. Among them, the single-degenerate \citep[SD,][]{1973ApJ...186.1007W, 1982ApJ...257..780N, 1982ApJ...253..798N, 2018SSRv..214...67N} and the double-degenerate \citep[DD,][]{1984ApJ...284..719I, 1984ApJ...277..355W, 2018ApJ...868...90T, 2019ApJ...885..103T} scenarios can explain a range of observed properties of SNe Ia. The WD accretes matter from a non-degenerate companion star (main-sequence, red-giant, He star) for the SD scenario. In the DD case, the explosion occurs when two white dwarfs merge. Another possible progenitor scenario is core-degenerate \citep[CD,][]{1974ApJ...188..149S, 2015MNRAS.450.1333S} which is a result of the merger of a white dwarf with an asymptotic giant branch (AGB) star. 

There have been a few studies to understand the progenitors of SNe Iax. A blue progenitor was detected for SN~2012Z in deep pre-explosion Hubble Space Telescope (HST) images (\citealt{2014Natur.512...54M, 2021arXiv210604602M}). The colors and luminosity indicated the progenitor to be a white dwarf accreting matter from a helium star. In the case of SN~2004cs and SN~2007J, He\, {\sc i} emission feature was detected in their post-maximum spectrum (\citealt{2009AJ....138..376F, 2013ApJ...767...57F}), which was explained as being due to a C-O white dwarf accreting matter from a He-donor, or as a result of interaction with circumstellar material \citep{2009AJ....138..376F}. However, in the case of SN~2007J,  the large helium content ($\sim$ 0.01 \(M_{\odot}\)) challenges the helium shell accretion scenario on a \(M_{\rm{ch}}\) white dwarf \citep{2019A&A...622A.102M}. A source consistent with the position of SN~2008ha was detected in the post-explosion HST image, which could be the progenitor white dwarf remnant after the explosion, or the companion star. This source is redder than the progenitor of SN~2012Z \citep{2014ApJ...792...29F}. \cite{2009Natur.459..674V} proposed weak explosions due to the core collapse of massive stars such as Wolf-Rayet stars as the progenitor of SN~2008ha. These stars, due to their high mass-loss rate, are hydrogen deficient. Most massive star progenitor scenarios were rejected as progenitors for SN~2008ge from the star formation rate of the host galaxy \citep{2010AJ....140.1321F}. Using pre-explosion HST images for SN~2014dt, \cite{2015ApJ...798L..37F} ruled out red giant or horizontal branch stars (\(M_{\rm initial}\) $\ge$ 8 \(M_{\odot}\)) and massive main-sequence stars (\(M_{\rm initial}\) $\ge$ 16 \(M_{\odot}\)) as progenitors. SN~2014dt shows mid-IR flux excess consistent with emission from newly formed dust. The derived mass-loss rate is consistent with either a red-giant or an asymptotic giant branch (AGB) star \citep{2016ApJ...816L..13F}.   

There is a range of explosion models proposed to explain the observed diversity of SNe Ia. The one-dimensional (1D) subsonic carbon deflagration in a Chandrasekhar mass (\(M_{\rm{ch}}\)) C-O white dwarf \citep{1984ApJ...286..644N} produces sufficient amount of $^{56}$Ni (0.5 - 0.6 \(M_{\odot}\)) and IME's to explain a range of normal SNe Ia. However, studies show that the deflagration turns into a supersonic detonation at a transition density (\citealt{1993A&A...270..223K, 1995ApJ...444..831H, 1996ApJ...457..500H, 2002ApJ...568..791H, 2013MNRAS.429.1156S, 2013MNRAS.436..333S}). By varying the transition density, a wide range of $^{56}$Ni mass can be produced. These are called deflagration-to-detonation transition (DDT). These models can reproduce the observed luminosity in normal and sub-luminous Ia and also the abundance stratification \citep{2005MNRAS.360.1231S} of the elements in the ejecta. Another variation of the standard detonation is the pulsational-delayed detonation (\citealt{1995ApJ...444..831H, 1996ApJ...457..500H, 2014MNRAS.441..532D}), in which deflagration causes expansion of the white dwarf followed by the infall of the expanding matter and hence compression of the white dwarf. This compression leads to a detonation at some particular density. It allows for more unburnt material in the ejecta. The widely favored model is the \(M_{\rm{ch}}\) explosion of C-O white dwarf \citep{2004MNRAS.350.1301H}. However, sub-Chandrasekhar mass detonation models can also reproduce a range of the observed properties of SNe Ia (\citealt{2010ApJ...719.1067K, 2012MNRAS.420.3003S, 2018ApJ...854...52S}).

For SNe Iax, the lower line velocities suggest that the explosion energies must be lower. The explosion produces less amount of ejecta as compared to SNe Ia and leaves behind a bound remnant \citep{2021PASJ..tmp...86K}. The abundance distribution in the ejecta is mixed (\citealt{2003PASP..115..453L, 2004PASP..116..903B}). These features can be explained by pure deflagration of \(M_{\rm{ch}}\) C-O white dwarf of varying strengths (\citealt{2013MNRAS.429.2287K, 2014MNRAS.438.1762F}). These models can produce a range of $^{56}$Ni mass and hence the luminosity observed in bright and intermediate luminosity SNe Iax. For the fainter SNe Iax, like SN~2008ha, pure deflagration in \(M_{\rm{ch}}\) carbon-oxygen-neon (C-O-Ne) white dwarfs has been proposed \citep{2015MNRAS.450.3045K}. 

SNe Iax show signatures of unburned carbon/oxygen in their spectra. These are essential in understanding the explosion models. Three-dimensional deflagration will produce unburned material in the inner parts of the ejecta near the center. A detonation will burn the materials in the inner regions and leave unburned material at lower density outer regions \citep{2003Sci...299...77G}. The velocity of the unburned layers can constrain the models. The presence of C-O indicates the nature of the progenitor - carbon-oxygen (C-O) white dwarfs \citep{2007PASP..119..360P} or carbon-oxygen-neon white dwarfs \citep{2015MNRAS.450.3045K} for the lower luminous subclass of SNe Iax.

The observed diversity and the possibility of a diverse class of progenitors make it important to study SNe Iax. The study of a recent SN Iax,  SN~2020csk is presented in this work. SN~2020sck was discovered by \cite{2020TNSTR2629....1F} on 2020 August 25, 10:03 UT (JD = 2459086.92) with a magnitude of 19.7~mag in ZTF-$r$ filter. The last non-detection was reported on 2020 August 25 09:07 UT with a limiting magnitude of 20.63~mag in the same filter. The object was classified as a SN Iax by \cite{2020TNSCR2685....1P} based on a spectrum obtained on 2020 August 30, 03:43 UT (JD=2459091.66) by the Liverpool Telescope. The important parameters of SN~2020sck and its host galaxy are presented in Table~\ref{tab:SN_params}. 

The details of the observations are presented in Section~\ref{Observations}. Section~\ref{Light Curve Analysis} and Section~\ref{Spectral Analysis} present the analysis of light curve and spectral evolution. Spectral modeling of SN~2020sck with \texttt{SYN++} and \texttt{TARDIS} is presented in Section~\ref{Spectral Modeling}. In Section~\ref{Host} we discuss the host galaxy and its properties. Possible explosion models are discussed in Section~\ref{Explosion Models}. Finally, we summarise our results in Section~\ref{Summary}.

\begin{figure}
\centering
\includegraphics[width=\columnwidth]{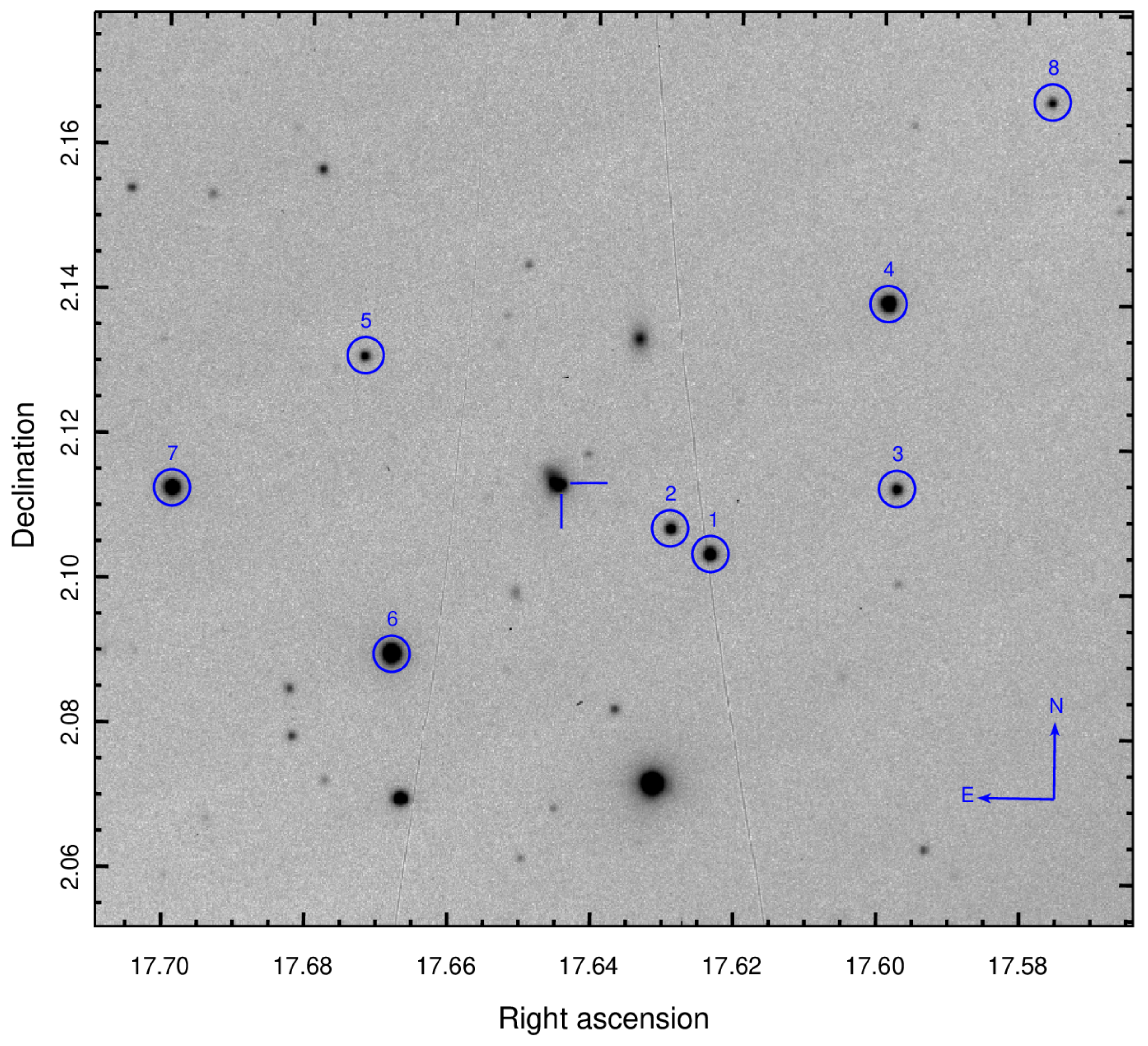}
\caption{The field of SN~2020sck. This is a $\sim$\,7\,$\times$\,7 arcmin$^2$ image in $B$-band (300 sec exposure) taken with HCT on 2020 September 11. The stars circled in blue are the secondary standards used for calibration purpose. The SN is marked with crosshairs.}
\label{Fig1}
\end{figure}

\begin{figure}
\centering
\includegraphics[width=\columnwidth]{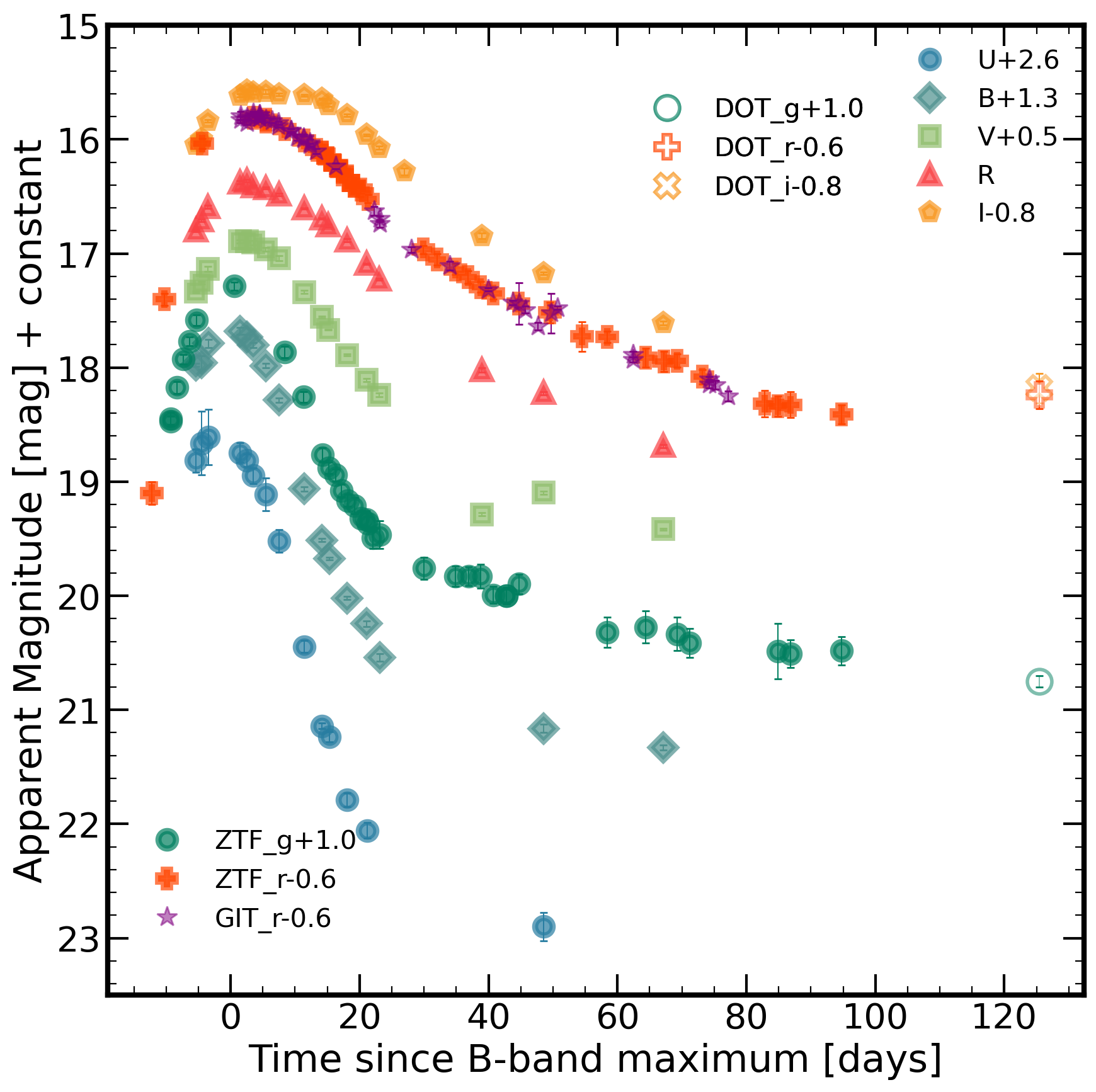}
\caption{$UBVRI$, ZTF-$g$, ZTF-$r$, GIT-\(r^{\prime}\) band light curves of SN~2020sck. Also plotted are the $g$, $r$ and $i$-band magnitudes obtained from DOT. The phase is measured with respect to $B$-band maximum. The light curves in individual bands have been shifted for representation purpose. The $UBVRI$ band magnitudes are in {\it{Vega}} system while the ZTF-$g$, ZTF-$r$, GIT-\(r^{\prime}\), DOT-$g$, DOT-$r$ and DOT-$i$ band magnitudes are in {\it{AB}} system. The HCT and GIT photometric data is tabulated in Table~\ref{tab:photlog}. The DOT photometry is listed in Table~\ref{tab:dot_photlog}.}
\label{Fig2}
\end{figure} 

\section{Observations and Data Reductions} \label{Observations} 

\subsection{Optical photometry}

Imaging of SN~2020sck in Bessell's \textit{UBVRI} bands was carried out with the Himalayan Faint Object Spectrograph Camera (HFOSC) mounted on the 2.0 m Himalayan Chandra Telescope located at the Indian Astronomical Observatory (IAO) at Hanle, India \footnote{\url{https://www.iiap.res.in/?q=telescope_iao}} (Fig.~\ref{Fig1}). Photometric observations with HCT started on 2020 August 31, at 5.4 days before $B$-band maximum and continued till 2020 November 13. A set of local standard stars in the SN field was calibrated using Landolt standards PG~1633+099 and PG~0231+051 observed on 2020 September 01, PG~2331+055, PG~2213-006 observed on 2020 September 07, PG~2331+055 observed on 2020 September 24, PG~0231+051 observed on 20 September 27 and PG~2213-006 observed on 2020 October 25. The $UBVRI$ magnitudes of the local standard stars are listed in Table~\ref{tab:standardlog}. To obtain the SN magnitudes, template subtraction has been performed. Deep stacked images of the SN field was observed on 2021 July 16 under good seeing conditions after the SN has faded beyond the detection limit. Details of the data reduction can be found in \cite{2021MNRAS.503..896D}.

SN~2020sck was followed up in SDSS-$r'$ filter with the 0.7~m fully robotic GROWTH-India Telescope (GIT) \footnote{Global Relay of Observatories Watching Transients Happen (\url{https://www.growth.caltech.edu/}), (\url{https://sites.google.com/view/growthindia/})} at IAO, equipped with a 2148 $\times$ 1472 pixels Apogee Camera. The observations began on 2020 September 07 and continued till 2020 November 23. GIT can be used in both targeted and tiled modes of operation. We used the targeted mode of operation and obtained 300 sec exposure images. PanSTARRS image of the field in $r$-filter was used as the reference image for host galaxy subtraction and the photometric zero points were calculated using the PanSTARRS catalog \citep{2018AAS...23143601F}. We used \texttt{PYZOGY} which is based on ZOGY algorithm \citep{2016ApJ...830...27Z} to perform the image subtraction. Finally, the PSF model generated by PSFex \citep{2011ASPC..442..435B} for the GIT image was used for photometry of the SN. The SN magnitudes in Bessell $UBVRI$ and SDSS-$r'$ are listed in Table~\ref{tab:photlog}.

Late phase observations of SN~2020sck was carried out with the ARIES-Devasthal Faint Object Spectrograph and Camera mounted on the axial port of the 3.6~m Devasthal Optical Telescope (DOT) \citep{2019BSRSL..88...31O}. Imaging in SDSS-$g$, $r$ and $i$ bands were performed on 2021 January 11. The data has been reduced in the standard manner as for HFOSC. To obtain the SN magnitudes, template subtraction has been performed with SDSS images in $g$, $r$ and $i$ bands. The SN magnitudes were calibrated using photometric zero points calculated using SDSS catalog \citep{2020ApJS..249....3A}. Table~\ref{tab:dot_photlog} lists the magnitudes obtained with DOT.

SN~2020sck was also observed with the Zwicky Transient Facility (ZTF, \citealp{2019PASP..131a8002B}). The photometric data in $g$ and $r$ bands was collected from the public archive\footnote{\url{https://alerce.online/}}.

\subsection{Optical Spectroscopy}

Spectroscopic monitoring of SN~2020sck with HCT started on 2020 October 31 (JD=2459093.31) and continued till 2020 October 03 (JD=2459126.34).
Low-medium resolution spectra were obtained using grisms Gr7 (3500-7800 \AA) and Gr8 (5200-9100 \AA) available with HFOSC. The log of spectroscopic observations is provided in Table~\ref{tab:speclog}. The spectra have been corrected for a redshift of z=0.016. Telluric features have been removed from the spectra. The data reduction was performed using the procedure described in \cite{2021MNRAS.503..896D}. 

\begin{figure*}
\centering
\includegraphics[width=0.7\linewidth]{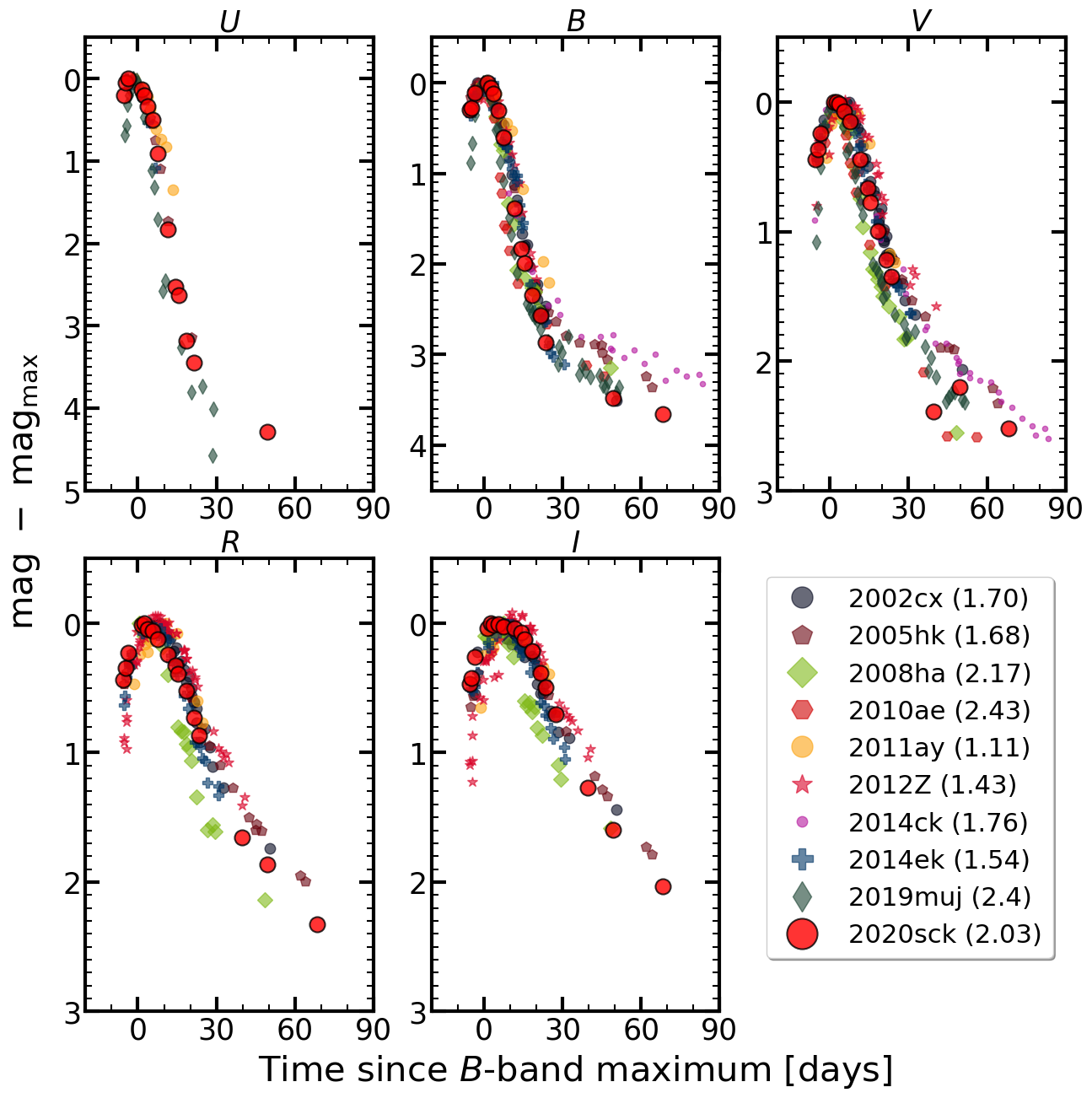}
\caption{$UBVRI$ light curves of SN~2020sck plotted along with other SNe Iax. The phase is measured with respect to $B$-band maximum. The light curves have been shifted to match with their respective peak magnitudes. The \(\Delta m_{15}(B)\) of each SN are quoted in parentheses.}
\label{Fig3}
\end{figure*} 

\begin{figure*}
\centering
\includegraphics[width=0.8\linewidth]{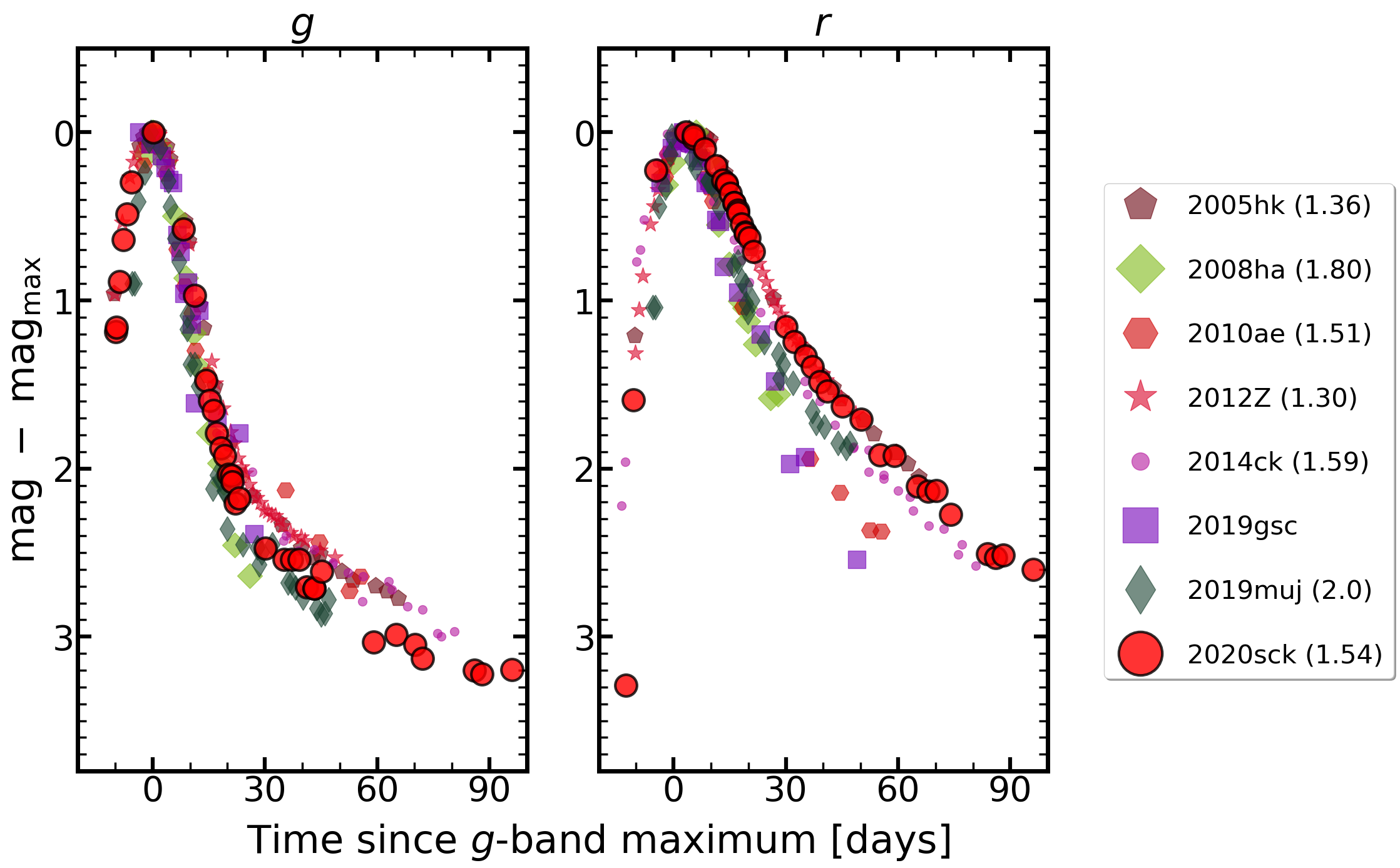}
\caption{ZTF-$g$ and ZTF-$r$ light curves of SN~2020sck plotted along with other SNe Iax in similar filters. The phase is measured with respect to $g$-band maximum. The light curves have been shifted to match with their respective peak magnitudes. The \(\Delta m_{15}(g)\) of each SN are quoted in parentheses. The light curve data for SN~2020sck has been obtained from {\url{https://alerce.online/object/ZTF20abwrcmq}}.}
\label{Fig4}
\end{figure*} 

\begin{deluxetable}{lcc}[htb]
\tablecaption{Parameters of SN 2020sck and its host galaxy.} \label{tab:SN_params}
\tablehead{\colhead{Parameters} & \colhead{Value} & \colhead{Ref.}}
\startdata
{\textit{SN 2020sck/ZTF20abwrcmq}:}  &               &          \\
\\
RA (J2000) 			    & $\alpha\,=\,01^{\rm h} 10^{\rm m} 34\fs84$ 	             & 2 \\
DEC (J2000)             & $\delta\,=\,+02\degr 06\arcmin 50\farcs15$ 	             & 2 \\
Discovery Date 		    & 2020 August 25 10:03 UT                             & 2 \\
			            & (JD 2459086.92)            			                     &   \\
Last non-detection      & 2020 August 25 09:07 UT		                         & 2 \\	            
                		& (JD 2459086.88)	                  		                 &   \\
Date of explosion       & 2020-08-20 21:15 UT                                 & 1 \\  
                        & 2459082.39$^{+1.57}_{-1.37}$                               &   \\
Date of $B$-band Maxima   & 2020 September 06 08:09 UT 	 & 1 \\
			            & (JD 2459098.84$\,\pm\,$0.30)    			                 & \\
$\Delta m$$_{15}$($B$)       & $2.03\pm0.05$ mag                                      & 1 \\	
Galaxy reddening        & $E(B-V)$\,=\,0.0256$\,\pm\,$0.0014 mag                     & 3 \\
Host reddening          & $E(B-V)$\,=\,0.00 mag                                      & 1 \\
$^{56}$Ni mass          & $M_{\rm Ni}$ $=$ $0.13^{+0.02}_{-0.01}$ \(M_\odot\)            & 1 \\
Ejected mass            & $M_{\rm ej}$ $=$ 0.34$^{+0.07}_{-0.10}$ \(M_\odot\)                     & 1 \\
Kinetic energy          & $E_{\rm K}$ $=$ 0.05$^{+0.01}_{0.01}$\(\times 10^{51}\) erg~s$^{-1}$        & 1 \\
\\
\hline
{\textit{Host galaxy}:} \\
\\
Name         		    & 2MASX J01103497+0206508                     \\
Type                    & H-II galaxy 						                & 4 \\
RA (J2000) 			    & $\alpha\,=\,01^{\rm h} 10^{\rm m} 34\fs99$ 	    & 4 \\
DEC (J2000)             & $\delta\,=\,+02\degr 06\arcmin 51\farcs48$ 	    & 4 \\
Redshift               & z\,=\,0.016$\,\pm\,$0.00010                   & 4 \\
Distance modulus        & $\mu$\,=\,34.24$\,\pm\,$0.22 mag 			                 & 5 \\ 
12 $+$ log($\rm \frac{O}{H}$) &  8.54$\pm$0.05 dex                              & 1 \\
SFR                         & 0.09 \(M_{\odot}\) yr\(^{-1}\)                & 1 \\
\enddata
  \tablecomments{(1) This paper;
(2) \citealt{2020TNSTR2629....1F, 2020TNSCR2685....1P}; \\
~~~~(3) \cite{2011ApJ...737..103S};
(4) \cite{2006AJ....131.1163S}; \\
(5) \url{http://leda.univ-lyon1.fr/}}
\end{deluxetable}

\subsection{Extinction and distance modulus}

The reddening due to the SN host galaxy for SNe Ia can be estimated from the $(B-V)$ color evolution of the SN during 30-90 days since the $B$ band maximum \citep{1999AJ....118.1766P}. However, this relation may not strictly hold for SNe Iax, due to the scatter in the evolution \citep{2013ApJ...767...57F}. The reddening within the host galaxy can also be estimated by the detection of interstellar Na\, {\sc i} D line (\citealt{2003fthp.conf..200T, 2012MNRAS.426.1465P}). We do not detect the Na\, {\sc i} D line in our low-resolution spectra.  Therefore, assuming zero host galaxy extinction, we correct the data only for the Galactic reddening of $E(B-V)$=0.0256 \citep{2011ApJ...737..103S} with $R_{V}$ = 3.1 \citep{1999PASP..111...63F}. From the prominent hydrogen emission lines in the spectra of SN~2020sck, we estimate a redshift of $z$=0.016 for the host galaxy. Using a distance modulus of $\mu$=$\rm 34.24\pm0.22$ mag  \footnote{\url{http://leda.univ-lyon1.fr/}} derived from the Virgo infall assuming $H_{0}$ = 70~km~s\(^{-1}\)~Mpc\(^{-1}\) \citep{2014A&A...570A..13M}, we find the absolute magnitude \(M_{\rm{B}}\) to be $-$17.81 $\pm$ 0.22 mag.

\section{Light curve} \label{Light Curve Analysis}

\begin{table*}
\centering
\caption{Photometric parameters of SN 2020sck.}
\label{tab:Phot_params}
\begin{tabular}{lcccccc}
\hline\hline
Filter &  $\rm \lambda_{eff} (\text \AA)$  &    JD (Max)       & $m_{\lambda}^\text{max}$     & $\Delta m_{15}(\lambda)$ & $M_{\lambda}^\text{max}$ & Colors at $B$ max\\
\hline\hline  
$U$ & 3663.6 &2459096.84 $\pm$ 0.57   & 16.03 $\pm$ 0.06             & 2.17 $\pm$ 0.06  & $-$18.33 $\pm$ 0.23   & $-$ \\
$B$ & 4363.2 &2459098.84 $\pm$ 0.30   & 16.53 $\pm$ 0.02             & 2.03 $\pm$ 0.05  & $-$17.81  $\pm$ 0.22  & $(U-B)_{\rm 0}$ = $-$0.35$\pm$0.02 \\ 
$V$ & 5445.8 &2459100.84 $\pm$ 0.26   & 16.41 $\pm$ 0.02             & 0.80 $\pm$ 0.03  & $-$17.91  $\pm$ 0.22  & $(B-V)_{\rm 0}$ = $-$0.08$\pm$0.03 \\
$R$ & 6414.2 &2459101.57 $\pm$ 0.48   & 16.39 $\pm$ 0.01             & 0.42 $\pm$ 0.02  & $-$17.93  $\pm$ 0.22  & $(V-R)_{\rm 0}$ = $-$0.01$\pm$0.01 \\
$I$ & 7978.8 &2459103.58 $\pm$ 1.40   & 16.37 $\pm$ 0.02             & 0.27 $\pm$ 0.02  & $-$17.91  $\pm$ 0.22  & $(R-I)_{\rm 0}$ = $-$0.02$\pm$0.01 \\
$ZTF-g$ & 4722.7 &2459099.14 $\pm$ 0.34   & 16.27 $\pm$ 0.03         & 1.54 $\pm$ 0.04  & $-$18.07  $\pm$ 0.22  & $-$ \\
$ZTF-r$ & 6339.6 &2459100.28 $\pm$ 0.76   & 16.34 $\pm$ 0.03         & 0.49 $\pm$ 0.03  & $-$17.96  $\pm$ 0.22  & $-$ \\
\hline
\end{tabular}
\end{table*}

\subsection{Light Curve Analysis}  \label{light curve}

The light curves of SN~2020sck in $U$, $B$, $g$, $V$, $r$, $R$, $I$ bands are shown in Fig.~\ref{Fig2}. SN~2020sck was followed from $-5.38$~d to $+67.11$~d since the $B$-band maximum in $UBVRI$ and $-13.36$~d to $+95.42$~d  since $r$-band maximum in ZTF-$g$ and ZTF-$r$ bands. We fit the $UBVRI$, ZTF-$g$ and ZTF-$r$ bands with Gaussian process regression \citep{3569} using the \texttt{gaussian\_process} package in \texttt{scikit-learn} \citep{scikit-learn} and find the epoch of maximum, maximum magnitude and the associated errors in each band. Table~\ref{tab:Phot_params} lists the important photometric parameters of SN~2020sck. SN~2020sck reached its peak $B$-band magnitude of 16.53 $\pm$ 0.02~mag at JD~2459098.84. The maximum in $U$-band occurred at $-$2.0~d and that in $V$, $R$ and $I$-bands at $+$2 days, $+2.8$ and $+$4.7 days, respectively since the $B$-band maximum. This indicates that the ejecta is cooling with time and follows a simple thermal model. The delay in $V$-band with respect to $B$-band maximum is similar to that seen in SN~2002cx and SN~2005hk. The $R$ and $I$-bands show no secondary maximum as are seen for SNe Ia. In Fig.~\ref{Fig3} the light curves of SN~2020sck in $UBVRI$ have been compared with other SNe Iax. SN~2020sck has a decline rate of $\Delta m_{15}(B)$=$2.03\pm0.05$ mag in $B$-band which is faster than bright SNe Iax like SN~2002cx and SN~2005hk and slower than some of the low luminosity objects like SN~2008ha, SN~2010ae, SN~2019muj. SN~2020sck shows a decline rate in $V$-band ($\Delta m_{15}(V)$=$0.80$ mag) similar to SN~2002cx and SN~2012Z. The redder bands show slower decline ($\Delta m_{15}(R)$=$0.42$ mag, $\Delta m_{15}(I)$=$0.27$ mag).

\begin{table*}
\centering
\setlength{\tabcolsep}{10pt}
\caption{Properties of the comparison sample.}
\label{tab:SNIax_sample}
\resizebox{\linewidth}{!}{%
\begin{tabular}{l c c c c c c c c}
\hline
SN   &  $M_{\rm B}$     &  $M_{\rm V}$  &  $\Delta m_{\rm 15}(B)$ & $\Delta m_{\rm 15}(V)$   & $\Delta m_{\rm 15}(g)$ & $\Delta m_{\rm 15}(r)$ &  12 + log($\rm \frac{O}{H}$) &  Reference \\
(Name)     &       (mag)         &    (mag) &   (mag)      & (mag)     & (mag)   & (mag)       & (dex)    &       \\
\hline\hline  
SN~2002cx &   -17.53$\,\pm\,$0.26    &  -17.49$\,\pm\,$0.22  & 1.70$\,\pm\,$0.1   & 0.73 &   --                & --   & --                 & 1, 2 \\
SN~2005hk &   -18.02$\,\pm\,$0.32    &  -18.08$\,\pm\,$0.29  & 1.68$\,\pm\,$0.05  & 0.92 &   1.36$\,\pm\,$0.01 & 0.70 & --                 & 2, 3 \\
SN~2008ha &   -13.74$\,\pm\,$0.15    &  -14.21$\,\pm\,$0.15  & 2.17$\,\pm\,$0.02  & 1.29 &   1.80$\,\pm\,$0.03 & 1.11 & 8.16$\,\pm\,$0.15  & 4 \\
SN~2009ku &   --                     &   --                  & -18.4              & --   &   0.59              & --   & --                 & 5  \\
SN~2010ae &  -13.44$\,\pm\,$0.54     &   -13.80              & 2.43$\,\pm\,$0.11  & 1.15 &   1.51$\,\pm\,$0.05 & 1.01 & 8.40$\,\pm\,$0.18  & 6  \\
SN~2011ay &  -18.15$\,\pm\,$0.17     &   -18.39$\,\pm\,$0.18 & 1.11$\,\pm\,$0.16  & 0.95 &   --                & --   &        --          & 7 \\
SN~2012Z  &  -17.61                  &   -18.04              & 1.57$\,\pm\,$0.07  & 0.89 &   1.30$\,\pm\,$0.01 & 0.66 & 8.51$\,\pm\,$0.31  & 8 \\
PS1-12bwh &   --                     &   --                  & --                 & --   &   1.35$\,\pm\,$0.09 & 0.60 & 8.87$\,\pm\,$0.19  & 9 \\
SN~2013en &   --                     &   --                  & --                 & --   &   --                & --   & --                 & 10 \\ 
SN~2014ck &  -17.37$\,\pm\,$0.15     &   -17.29$\,\pm\,$0.15 & 1.76$\,\pm\,$0.15  & 0.88 &   1.59$\,\pm\,$0.1  & 0.58 & --                 & 11 \\  
SN~2014dt &  -18.13$\,\pm\,$0.04     &   -18.33$\,\pm\,$0.02 & 1.35$\,\pm\,$0.06  & --   &   --                & --   & --                 & 12  \\
SN~2014ek &  -17.32$\,\pm\,$0.23     &   -17.66$\,\pm\,$0.20 & 1.54$\,\pm\,$0.17  & 0.90 &   --                & --   & --                 & 13  \\
SN~2015H  &   --                     &    --                 & --                 & --   &   --                & 0.69 & --                 & 14 \\
SN~2019gsc &  --                     &    --                 & --                 & --   &   --                & 0.91 & 8.10$\,\pm\,$0.06  & 15 \\
SN~2019muj & -16.36$\,\pm\,$0.06     &   -16.42$\,\pm\,$0.0  & 2.4                & 1.2  &   2.0               & 1.0  &   --               & 16 \\
\hline
\end{tabular}}
\newline \newline
References:\
(1) \citet{2003PASP..115..453L}; (2) \citet{2007PASP..119..360P} (3) \citet{2008ApJ...680..580S}; (4) \citet{2009AJ....138..376F}; (5) \citet{2011ApJ...731L..11N}; (6) \citet{2014A&A...561A.146S}; (7) \citet{2015MNRAS.453.2103S}; (8) \citet{2015ApJ...806..191Y}; 
(9) \citet{2017A&A...601A..62M}; (10) \citet{2015MNRAS.452..838L}; (11) \citet{2016MNRAS.459.1018T}; (12) \citet{2018MNRAS.474.2551S}; (13) \citet{2018MNRAS.478.4575L}; (14) \citet{2016A&A...589A..89M}; (15) \citet{2020ApJ...892L..24S}; (16) \citet{2021MNRAS.501.1078B}
\end{table*}

Fig.~\ref{Fig4} shows the comparison of ZTF-$g$ and ZTF-$r$ band light curves of SN~2020sck along with other SNe Iax in similar filters. The decline rate in $g$-band ($\Delta m_{15}(g)$=$1.54$ mag) is similar to SN~2010ae and SN~2014ck. SN~2020sck has the slowest decline in $r$-band with a $m_{15}(r)$=$0.49\pm0.03$ mag. The decline rate for SN~2005hk and SN~2012Z in $r$-band are 0.70~mag and 0.66~mag respectively. For the fainter SNe Iax, the decline rate in $r$-band is faster. Table~\ref{tab:SNIax_sample} provides the observed properties of the other SNe Iax used for comparison.

\begin{figure}
\centering
\includegraphics[width=\columnwidth]{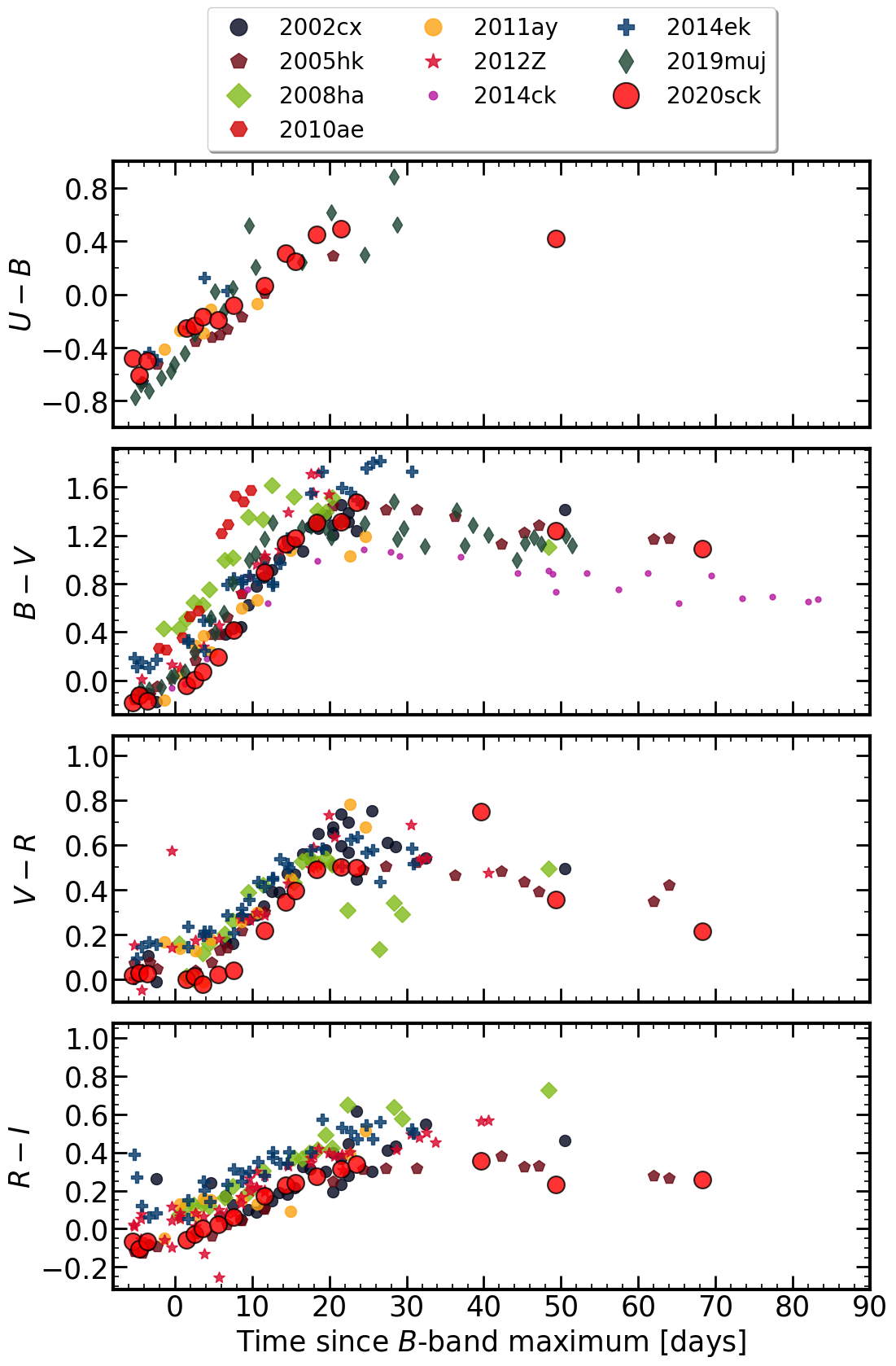}
\caption{$(U-B)$, $(B-V)$, $(V-R)$ and $(R-I)$ color evolution of SN~2020sck plotted with other SNe Iax.}
\label{Fig5}
\end{figure}

\begin{figure}
\centering
\includegraphics[width=\columnwidth]{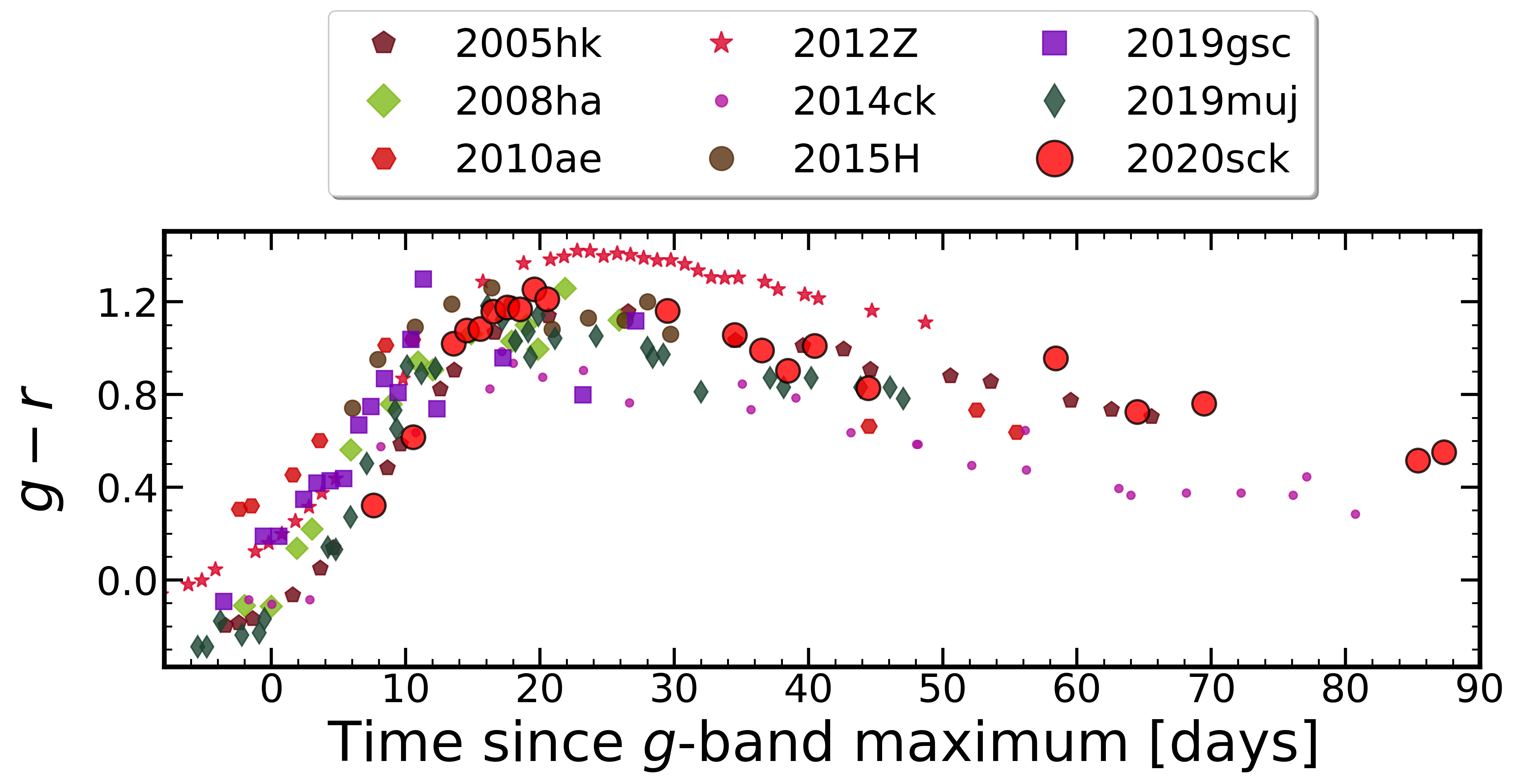}
\caption{$(g-r)$ color evolution of SN~2020sck plotted along with other SNe Iax.}
\label{Fig6}
\end{figure}

The light curve decline rate of SN~2020sck is similar to the lower luminosity SNe Iax in the blue bands, while it is similar to the brighter objects in the red bands. The $(U-B)$, $(B-V)$, $(V-R)$, $(R-I)$ and $(g-r)$ color evolution of SN~2020sck are plotted in Fig.~\ref{Fig5} and Fig.~\ref{Fig6} and compared with other SNe Iax. The overall trend of the color evolution of SN~2020sck is similar to other well studied SN Iax events. The $(U-B)$ color evolution is similar to SN~2005hk and SN~2011ay. The $(B-V)$ color is bluer near maximum in $B$-band ($-$0.08$\pm$0.03 mag) and follows the same trend as other SNe Iax in the later phase. In comparison, the $(B-V)$ color at $B$-max is 0.04 mag for SN~2002cx and $-$0.03 mag for SN~2005hk. The $(V-R)$ color is also bluer than the comparison SNe. The $(R-I)$ and $(g-r)$ color evolution is similar to SN~2005hk.

\subsection{Estimation of time of first light} \label{sec: expl_time}
\begin{figure}
\centering
\includegraphics[width=\columnwidth]{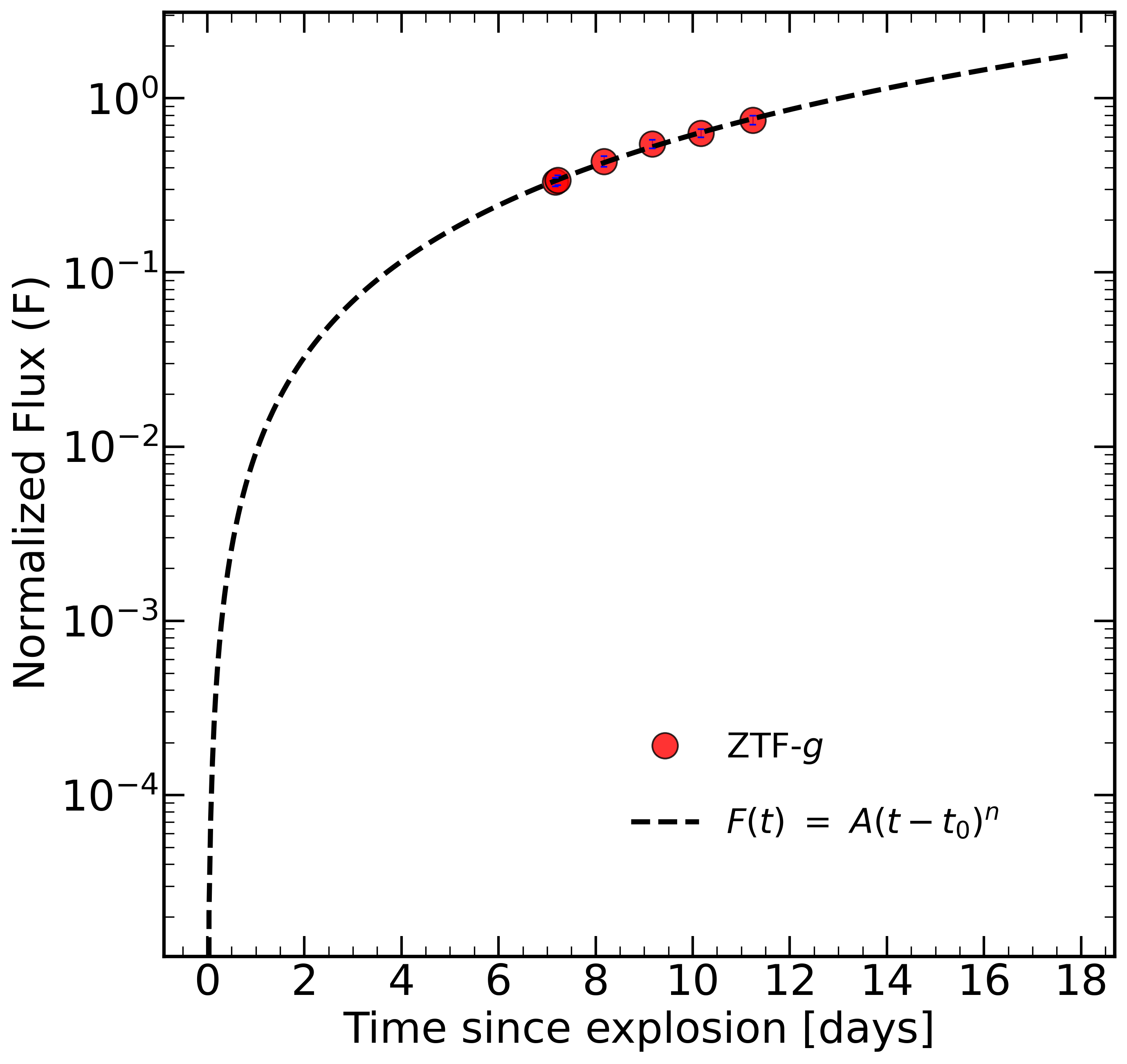}
\caption{$F \sim$ $t^{n}$ fit to the pre-maximum ZTF-$g$ band light curve. Flux error is shown in blue.}
\label{Fig7}
\end{figure}

During the early times of the explosion, the luminosity is proportional to the surface area of an expanding fireball and hence increases as $t^{2}$, where $t$ is the time since the explosion. This assumes that the photospheric velocity and temperature do not change significantly during this phase \citep{1999AJ....118.2675R}. SN~2020sck was monitored with ZTF soon after its discovery ($\sim$~JD~2459086) in $g$ and $r$-bands. This allows us to place a constraint on the time of first light. We fit the early $g$-band data of ZTF with a power law of the form 
\begin{equation}
\label{eq:tfl}
    F(t) = A(t-t_{0})^{n}
\end{equation}
where $A$ is a normalization constant, $t_{\rm 0}$ is the time of first light and $n$ is the power-law index. For the ``fireball model'' the value of $n$ is 2. The variation from this value hints towards the distribution of $^{56}$Ni in the ejecta, with a lower index pointing towards higher degrees of mixing \citep{2015MNRAS.446.3895F}. The value of $n$ varies from $\sim$ 1.5 to $\sim$ 3.5. In the fit, we kept $n$ as a free parameter. We aimed to fit the $g$-band flux with a starting value of $t_{\rm 0}$ = 2459086 from the non-detection. However, from the fit, we get an unrealistic value of $n$ = 0.39. 
Next, we kept the starting value of $t_{\rm 0}$ between 2459080 and 2459087 and from the fit we obtain an explosion date of 2020 August 20 21:15 UT (JD = $2459082.39^{+1.57}_{-1.37}$) and an exponent(n) of $1.79^{+0.29}_{-0.33}$. We use JD 2459082.4 as the explosion date throughout the work. The power-law fit is shown in Fig.~\ref{Fig7}. From the fit, we estimate the rise time to the maximum in $g$-band as 16.75 days and in $r$-band as 17.89 days. The rise time for SN~2020sck is similarly to SN~2002cx-like objects, for which the rise time is $\sim$ 15.0 days. The rise times for SN~2005hk \citep{2007PASP..119..360P} and SN~2015H \citep{2016A&A...589A..89M} are 15.0 days and 15.9 days ($r$-band) respectively. While SN~2008ha \citep{2009AJ....138..376F}, SN~2012Z \citep{2015ApJ...806..191Y}, SN~2019muj \citep{2021MNRAS.501.1078B} has lower rise times of 10, 12.0 and 9.6 days respectively, the rise time for SN~2009ku \citep{2011ApJ...731L..11N} is 18.2 days close to that for SNe Ia ($\sim$ 19.0 days). 

\subsection{Estimation of nickel mass} \label{Ni_mass}

\begin{figure}
\centering
\includegraphics[width=\columnwidth]{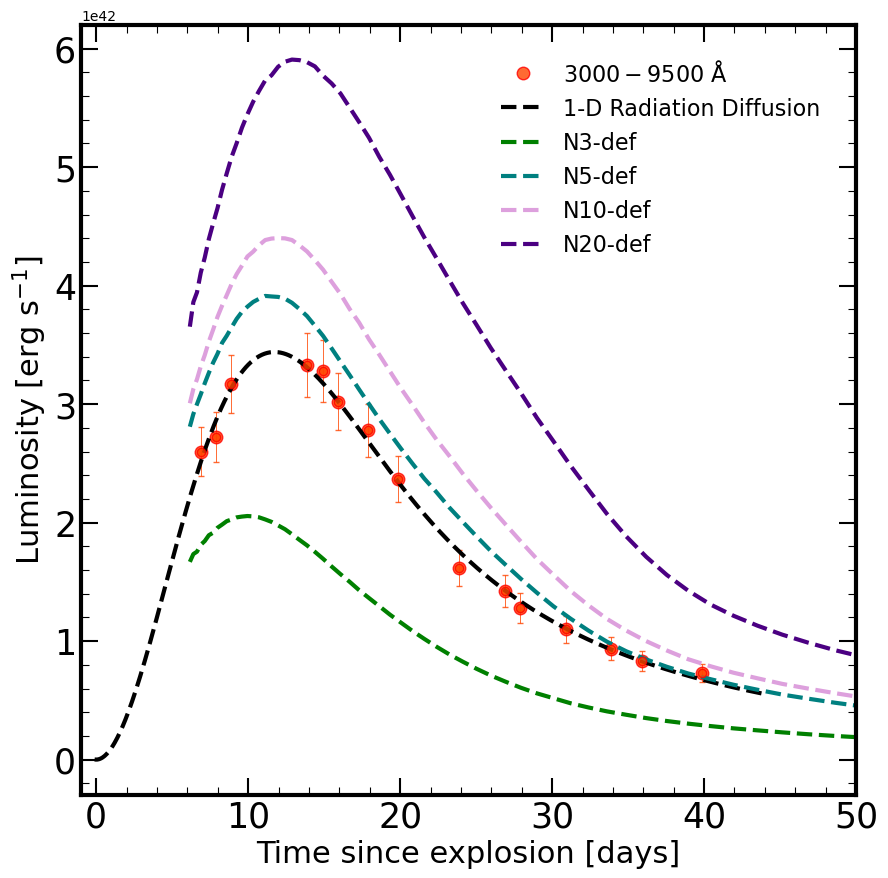}
\caption{The quasi-bolometric light curve of SN~2020sck fitted with 1D radiation diffusion model. Also plotted are the angle-averaged bolometric light curve from the three dimensional pure deflagration models of \(M_{\rm{ch}}\) WD \citep{2014MNRAS.438.1762F}. The models have been obtained from the Heidelberg Supernova Model Archive ({\sc hesma}).}
\label{Fig8}
\end{figure}

The bolometric light curve has been calculated using the $U$, $B$, $V$, $R$ and $I$-band magnitudes. The apparent magnitudes were corrected for the Milky Way reddening of $E(B-V)$=0.0256 and $R_{\rm V}$=3.1. The reddening corrected magnitudes were converted into flux units using zero points from \cite{1998A&A...333..231B}. A third-order spline curve was fit to the spectral energy distribution (SED) and the area under the curve was calculated using trapezoidal rule integrating from 3000 \AA~to 9500 \AA. For SNe Iax, due to the scatter in the light curve evolution, a well-defined correction factor in UV and IR does not exist. However, some SNe have been possible to observe in UV to IR wavelength range. For SN~2005hk, \cite{2007PASP..119..360P} have ignored the NIR flux contribution to the UVOIR bolometric light curve during the early phase and have used about $\sim 20\%$ contribution to the flux in UV before maximum. For SN~2012Z, \cite{2015ApJ...806..191Y} have shown that the ratio of the flux in IR to the combined flux in optical and IR increases from 0.15 to 0.3 from around 8 to 25 days since the explosion. However, the evolution is significantly different from that found in SN Ia. For SN~2014ck, \cite{2016MNRAS.459.1018T} have assumed a $10\%$ contribution to the UV flux at maximum. SN~2014dt showed a significant increase in NIR and mid-IR flux from about 100 days post-maximum in $B$-band \citep{2016ApJ...816L..13F}. For SN~2019gsc, \cite{2020MNRAS.496.1132T} found that the peak $gri$ bolometric luminosity is 53 $\%$ of the peak OIR bolometric luminosity.
To find the missing flux in UV and IR, a blackbody fit to the SED has been performed and added to the optical flux. This approach does not take into account the line-blanketing effects in the UV range and assumes that there is a contribution of UV and IR flux throughout the evolution of the bolometric light curve. The total flux thus obtained has been converted to luminosity assuming a distance modulus of $\mu$=$\rm 34.24$ mag. The quasi-bolometric light curve is shown in Fig.~\ref{Fig8}. We model the quasi-bolometric light curve as Gaussian process using the \texttt{gaussian\_process} package in \texttt{Scikit-learn} and estimate the peak luminosity for SN~2020sck to be \(L^{\rm{quasi-bol}}_{\rm{peak}} = (3.41\pm 0.25) \times10^{42}\) erg s$^{-1}$. The peak luminosity for the blackbody bolometric light curve is \(L^{\rm{BB}}_{\rm{peak}} = (5.51\pm 0.54) \times10^{42}\) erg s$^{-1}$. The peak quasi-bolometric luminosity is 62 $\%$ of the peak blackbody bolometric luminosity ($L_{\rm opt}$/ $L_{\rm BB}$). For SN~2019gsc, $L_{\rm opt}$/ $L_{\rm BB}$ is 69 $\%$ using a similar approach \citep{2020ApJ...892L..24S}.

To estimate the amount of nickel synthesized in the explosion, we fit the bolometric light curves with a modified radiation diffusion model (\citealt{1982ApJ...253..785A, 2008MNRAS.383.1485V, 2012ApJ...746..121C}). The modified model takes into account the diffusion of radioactive decay energy from $^{56}$Ni and $^{56}$Co and also the gamma-ray leakage from the ejecta. The output luminosity is expressed as:

\begin{equation}
\begin{split}
\label{eq:Arnett}
    L(t) = M_{\rm{Ni}} \mathrm{e}^{-x^{2}} [(\epsilon_{\rm{Ni}}-\epsilon_{\rm{Co}})\int_0^x 2 z \mathrm{e}^{z^{2}-2zy}\,\mathrm{d}z\\
         + \epsilon_{\rm{Co}}\int_0^x 2 z \mathrm{e}^{z^{2}-2yz+2zs}\,\mathrm{d}z]( 1 - \mathrm{e}^{-{(\frac{t_{\gamma}}{t}})^{2}}) 
\end{split}         
\end{equation}

where $x$ $\equiv$ $t$/$t$\(_{\rm{lc}}\), $t$ is the time since explosion (days) and \(t_{\rm{lc}}\) is the light curve time scale (days). $y$ $\equiv$ \(t_{\rm{lc}}\)/(2\(t_{\rm{Ni}}\)) with \(t_{\rm{Ni}}\) = 8.8~d, $s$ $\equiv$ [\(t_{\rm{lc}}\)(\(t_{\rm{Co}}\) - \(t_{\rm{Ni}}\))/(2\(t_{\rm{Co}}\)\(t_{\rm{Ni}}\))] with \(t_{\rm{Co}}\) = 111.3~d. \(M_{\rm{Ni}}\) is the initial Ni mass and \(t_{\gamma}\) is the gamma ray time scale (days). Large \(t_{\rm{\gamma}}\) means all the gamma rays and positrons are trapped. \(\rm \epsilon_{Ni} = 3.9 \times 10^{10}\ erg\ s^{-1}\ g^{-1}\) and \(\rm \epsilon_{Co} = 6.8 \times 10^{9}\ erg\ s^{-1}\ g^{-1}\) are the energy generation rates due to the decay of Ni and Co respectively. The fit parameters of the model are $t_{exp}$ - the epoch of explosion, $M_{\rm{Ni}}$ - the initial $^{56}$Ni mass produced, $t_{\rm{lc}}$ -  the light curve time scale and $t_\gamma$ - the gamma-ray leaking time scale. We can obtain the ejecta mass (\(M_{\rm{ej}}\)) and kinetic energy (\(E_{\rm{K}}\)) using the relations -

\begin{equation}
\label{eq:EjectaMass}
    M_{\rm{ej}} = 0.5 \frac{\beta c}{\kappa} v_{exp}t_{lc}^{2} 
\end{equation}

\begin{equation}
\label{eq:KineticEnergy}
    E_{\rm{K}} = 0.3 M_{ej} v_{exp}^{2}
\end{equation}

Here, $\beta$ = 13.8 is a constant of integration. $c$ is the speed of light. $v_{exp}$ is the expansion velocity of the ejecta.

To fit the model and find the model parmeters that best describe our quasi-bolometric light curve, we sampled the posterior disribution and maximized the posterior by maximizing the product of the likelihood and the prior. The likelihood function is -

\begin{equation}
    Log-likelihood = -\frac{1}{2}\sum_{i} (\frac{y_{model} - y_{i}}{y_{err}})^{2}
\end{equation}

\begin{figure}
\centering
\includegraphics[width=\columnwidth]{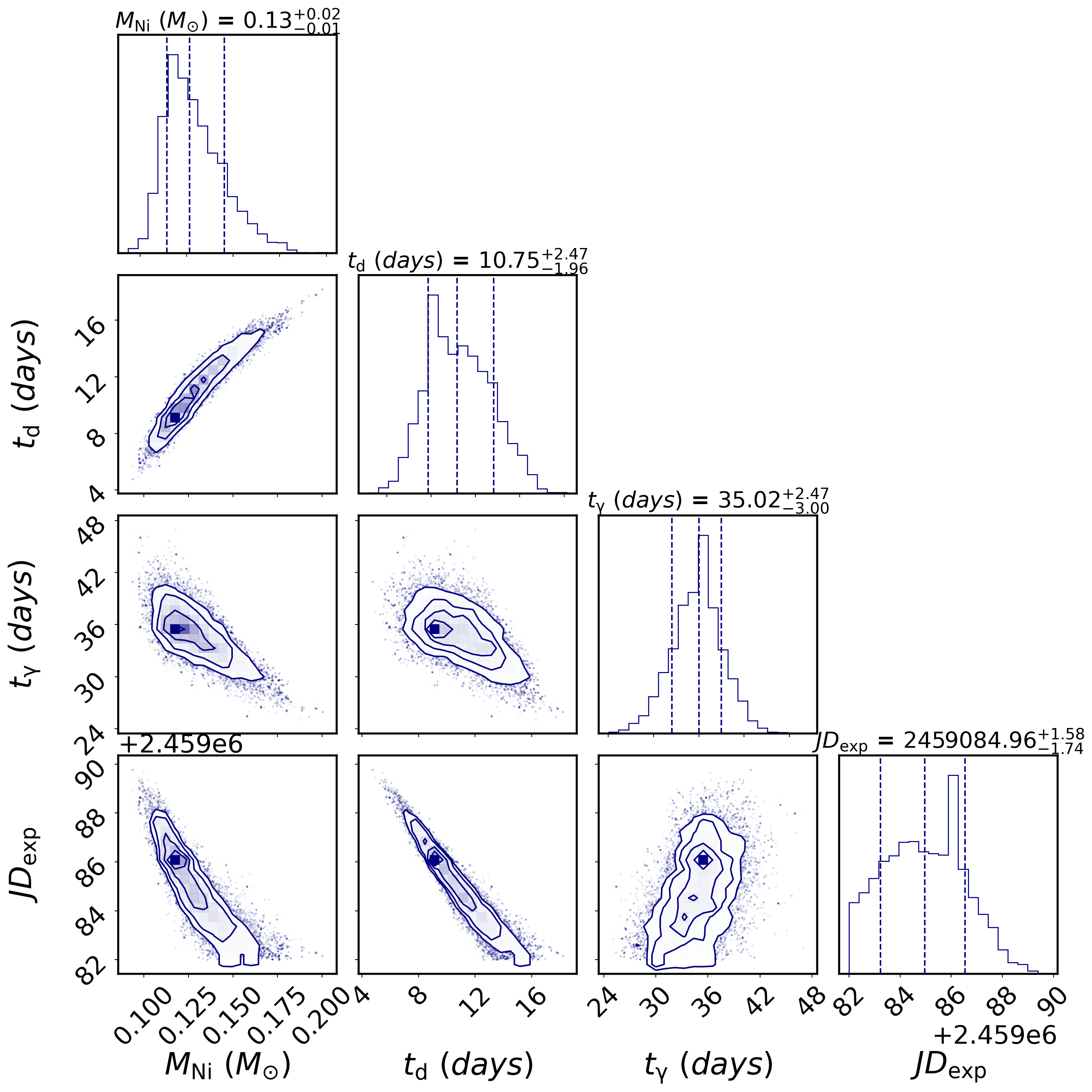}
\caption{One and two dimensional projections of the posterior probability distribution of the fit parameters of Eqn.~\ref{eq:Arnett} to the quasi-bolometric light curve of SN~2020sck. The vertical dashed lines show the 16th, 50th and 84th percentiles of the samples in the distribution. See Fig.~\ref{Fig8} for the fit.}
\label{Fig9}
\end{figure}

Here, $y_{model}$ and $y_{i}$ are the model luminosity and the measured luminosity respectively. $y_{err}$ is the error in the measured luminosity. The sum runs over all the datapoints. We used flat or uniform prior for the model parameters - 0 $<$ $M_{\rm Ni}$ $<$ 1.4 $M_{\odot}$, $t_{\rm lc}$ $>$ 0 d, $t_\gamma$ $>$ 0 d and 2459082 $<$ $t_{\rm exp}$ $<$ 2459090. We used the \texttt{emcee} package in \texttt{python} to find the posterior distribution of the model parameters \citep{2013ascl.soft03002F}. Fig.~\ref{Fig9} shows the one and two dimensional projections of the posterior distribution of the fit parameters.

The fit to the quasi-bolometric light curve gives $t_{exp}$ = $\rm 2459084.96^{+1.58}_{-1.74}$, $M_{\rm{Ni}}$ =  $0.13^{+0.02}_{-0.01}$ \(M_\odot\), $t_{\rm{lc}}$ = $10.75^{+2.47}_{-1.96}$ days and $t_\gamma$ = $35.02^{+2.47}_{-3.00}$ days. Using a constant optical opacity $\kappa_{\rm opt}$ = 0.1 cm$^{2}$g$^{-1}$ for a Fe dominated ejecta (\citealt{2000ApJ...530..757P, 2015MNRAS.453.2103S, 2020ApJ...892L..24S}) and an expansion velocity $v_{\rm exp}$ = 5,000 km s$^{-1}$ derived from the \texttt{SYN++} fitting of the near maximum spectrum, we get $M_{\rm{ej}}$ = $0.34^{+0.07}_{-0.10}$ \(M_\odot\) and a kinetic energy of explosion $E_{\rm{kinetic}}$ = $0.05^{+0.01}_{-0.01} \times 10^{51}$ erg. If we assume explosion of a \(M_{\rm{ch}}\) white dwarf, the bound remnant mass is 1.06 \(M_\odot\). The fit to the blackbody bolometric light curve gives \(M_{\rm{Ni}}\)=  $0.17^{+0.01}_{-0.01}$ \(M_\odot\).

The quasi-bolometric light curve of SN~2020sck has been compared with angle-averaged bolometric light curve from three-dimensional pure deflagrations of \(M_{\rm{ch}}\) carbon-oxygen white dwarfs. In these models, no delayed detonations occur to completely unbind the white dwarf and thus a bound remnant is left behind \citep{2014MNRAS.438.1762F}. The explosion is parametrized by multiple spherical ignition spots that burn simultaneously. This allows exploring a wide range of explosion strengths. The models N1, N3, N5, N10 and N20 corresponds to 1, 3, 5, 10 and 20 ignition spots respectively placed randomly around the centre of the white dwarf. The energy released in the explosion and the luminosity increases with an increasing number of spots. As the number of ignition spots increases, more matter is burnt and hence leads to higher expansion velocity of the ejecta. The model N5-def with $\Delta m_{15}(B)$=$1.69$ mag and $M_{\rm B}^\text{max}$ = $-$17.85 mag matches closely with SN~2020sck, which has a $\Delta m_{15}(B)$=$2.03$ mag and $M_{B}^\text{\rm max}$ = $-$17.81 mag.  In the N5-def model, the mass of $^{56}$Ni is 0.16 \(M_\odot\) , the ejecta mass is 0.372 \(M_\odot\), the mass of the bound remnant is 1.03 \(M_\odot\). These values match closely with that estimated for SN~2020sck from the quasi-bolometric light curve fit with the radiation diffusion model. The kinetic energy estimated by the N5-def model is 0.135 $\rm \times  10^{51}$ erg, the radiation diffusion model gives an estimate of 0.05 $\rm \times  10^{51}$ erg. The models with lesser number of ignition points (1, 3, 5) evolve in an asymmetric way compared to models with larger number of ingition kernels (150, 300 etc). So, moderate viewing angle dependence is possible in these deflagration models \citep{2014MNRAS.438.1762F}. The lower kinetic energy estimated by the radiation diffusion model can be explained if we assume that the explosion is similar to N5-def but with a lower line-of-sight velocity.

\begin{figure*}
\centering
\includegraphics[width=\textwidth]{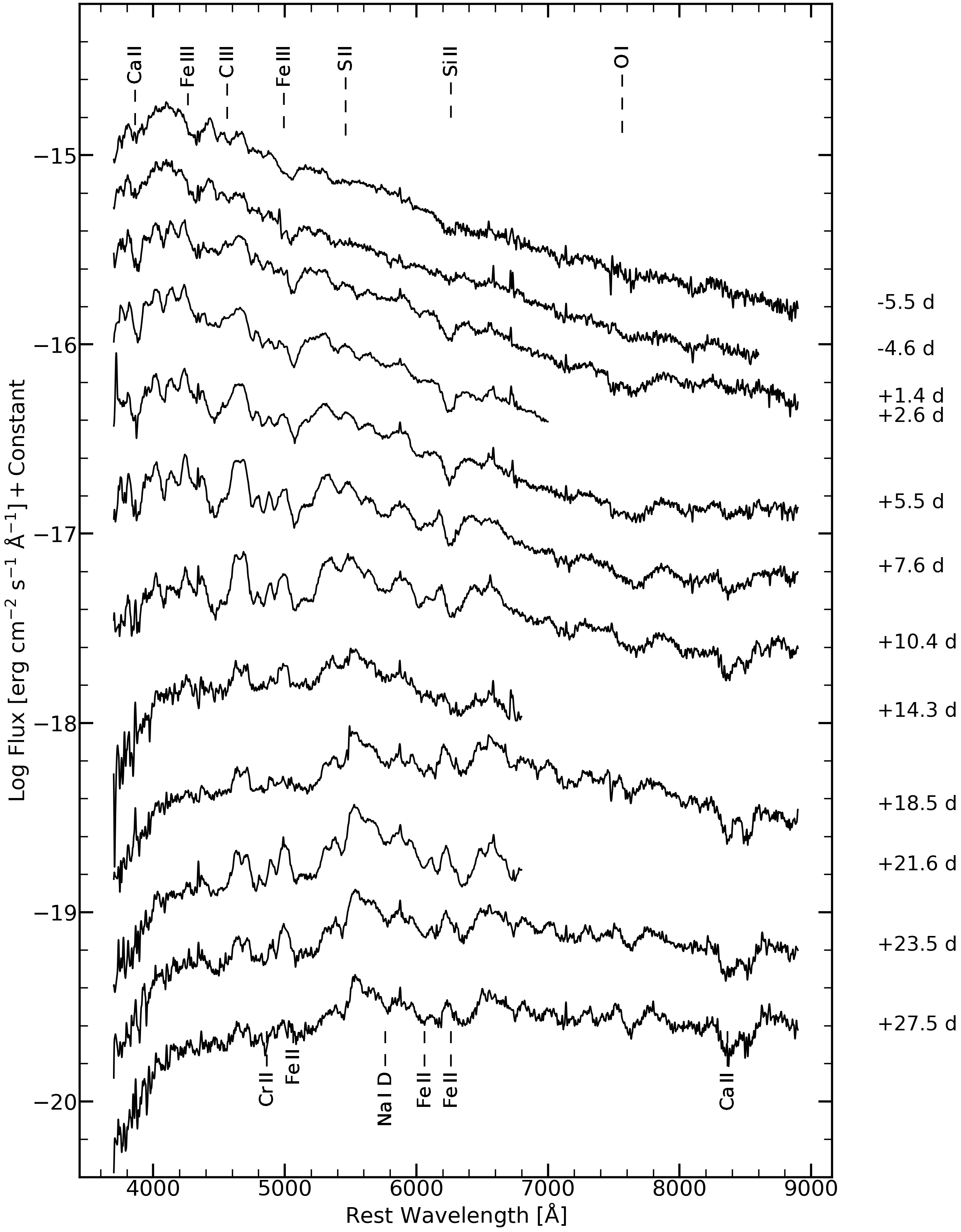}
\caption{Spectral evolution of SN~2020sck from $-$5.5 to $+$27.5 day since the $B$-band maximum. The spectra are dereddened and redshift corrected. The telluric features have been removed. Spectra have been smoothed for visual clarity. A detailed spectral line identification is shown in Fig.~\ref{Fig11}.}
\label{Fig10}
\end{figure*} 

\section{Spectral Analysis} \label{Spectral Analysis}

\subsection{Spectral evolution}
The spectroscopic evolution of SN~2020sck is shown in Fig.~\ref{Fig10}. The line identification has been done by comparing with SN~2002cx \citep{2004PASP..116..903B} and SN~2005hk \citep{2008ApJ...680..580S} around similar phases, and also with the spectrum synthesis code \texttt{SYN++} \citep{2011PASP..123..237T}. The spectra in the pre-maximum phase show a blue continuum and presence of absorption features due to Fe\,{\sc iii} ($\lambda$4420, 5075, 5156), Fe\,{\sc ii} ($\lambda$4924), Si\,{\sc iii} ($\lambda$4568), weak absorption feature of Co\,{\sc ii} ($\lambda$4161) around $\sim$4000 \AA, S\,{\sc ii} ($\lambda$5449, 5623) and an asymmetric weak absorption feature at $\sim$6200 \AA\ due to Si\,{\sc ii} ($\lambda$6355). We compare the spectra of SN~2020sck in the pre-maximum phase with other SNe Iax in panel (a) of Fig.~\ref{Fig10}. All the SNe except SN~2014ck show a blue continuum. The lines due to Fe\,{\sc iii} ($\sim$4420 \AA) and Fe\,{\sc iii} ($\sim$5156 \AA) are prominent in all the SNe with varying optical depth. The $-$5.4~d spectrum of SN~2020sck is similar to SN~2005hk and SN~2019muj. The weak IME features seen in SN~2020sck is possibly due to lower density and lesser optical depth in the outer regions, which allows us to probe the hotter inner regions of the ejecta. The presence of higher ionization states of IME's (Si\,{\sc iii}) and IGE's (Fe\,{\sc iii} ) also indicates a hot photosphere.
SN~2014ck shows deeper Si\,{\sc ii} ($\lambda$6355) and S\,{\sc ii} features. This is because of the lower luminosity and lower photospheric temperature. Prominent C\,{\sc ii} ($\lambda$6580) absorption feature is present in the spectrum of SN~2014ck and SN~2019muj in the pre-maximum phase. But for SN~2020sck, C\,{\sc ii} feature is not seen to be developed. This hints towards the fact that the outer layers of the ejecta has lesser C in SN~2020sck. 

\begin{figure*}
\centering
\includegraphics[width=0.9\linewidth]{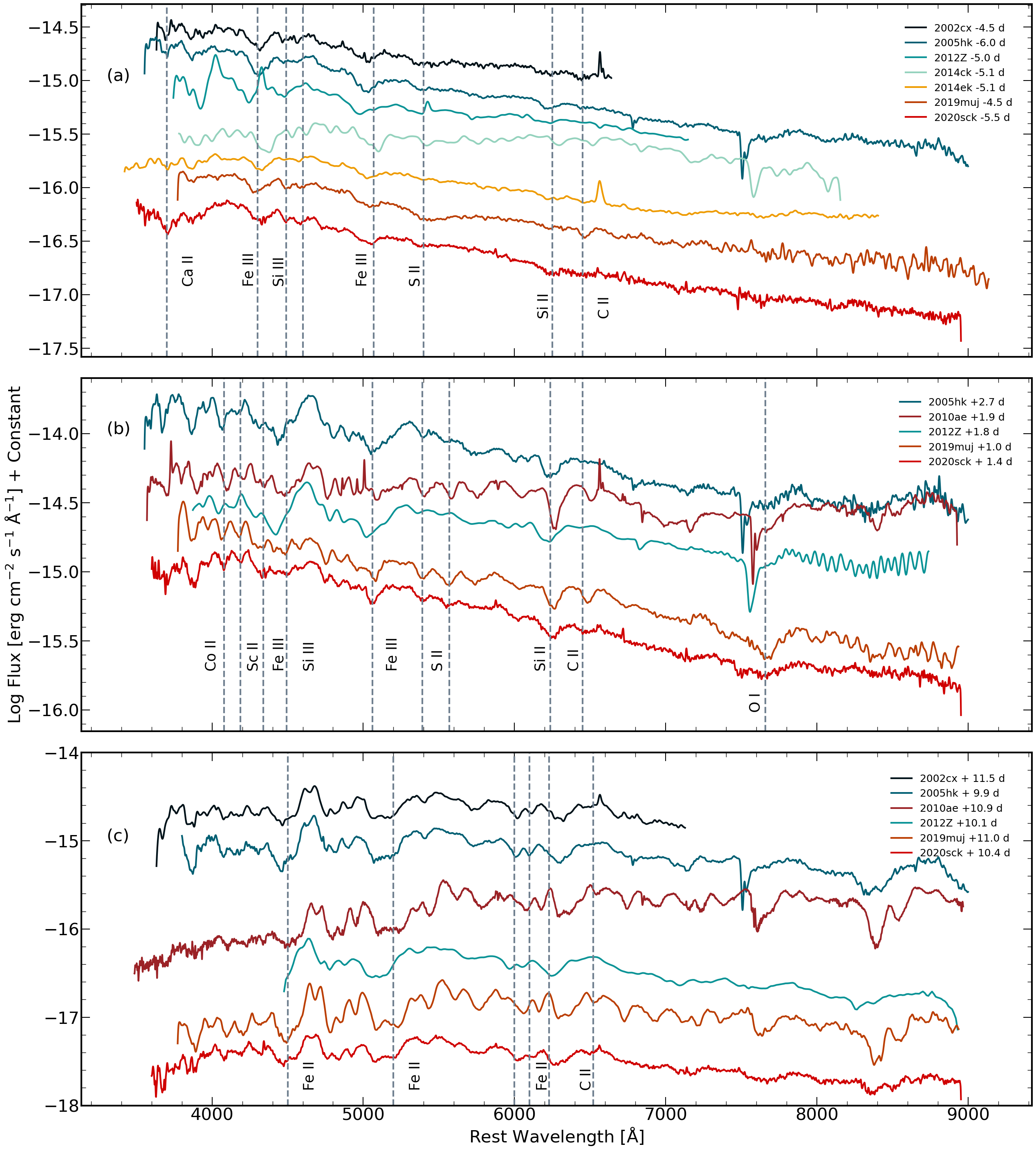}
\caption{Comparison of the spectra of SN~2020sck with other SNe Iax around similar phase. The dashed vertical lines show the position of the absorption minima of the ions for SN~2020sck. All the spectra have been smoothed for visual clarity.}
\label{Fig11}
\end{figure*}

Around maximum the absorption features of Si\,{\sc ii} ($\lambda$6355),  S\,{\sc ii} ($\lambda$5449, 5623) become prominent. Ca\,{\sc ii} H \& K ($\lambda\lambda$3934, 3968) and Ca\,{\sc ii} NIR triplet ($\lambda$8498) are seen to be developing. O\,{\sc i} ($\lambda$7775) feature is prominently visible. C\,{\sc ii} ($\lambda$6580) and C\,{\sc iii} ($\lambda$4647) absorption features can be seen in the spectrum taken at $+$1.4 day. C\,{\sc ii} ($\lambda$6580) feature begins to appear around maximum with a pseudo-equivalent width (pEW) of 5.25 $\pm$ 1.05 \AA\ at $+$1.4 d. This feature is present in our spectrum till $+$10.4 d. Appearance of C in the near-maximum phase implies that the C layer is mixed in the ejecta.
Comparing with other SNe Iax, it is seen that SN~2014ck and SN~2019muj also show prominent C\,{\sc ii} ($\lambda$6580) feature with pEW of 4 \AA\ and 12 \AA\ respectively. In the near-maximum phase, the spectrum is similar to SN~2019muj. The line profiles indicate lower velocities in SN 2020sck in comparison with SN~2005hk and SN~2012Z. The comparison of SN~2020sck with other SNe Iax around maximum is shown in panel (b) of Fig.~\ref{Fig11}.

Post-maximum, the Si\,{\sc ii} ($\lambda$6355) gets weakened and Fe\,{\sc ii} lines dominate (panel (c) of Fig.~\ref{Fig10}). The opacity of the Fe\,{\sc iii} lines decrease, or Fe\,{\sc iii} evolves to Fe\,{\sc ii} due to decrease in temperature. 
Na\,{\sc i} D absorption line can be seen to have developed. By $\sim$ 2 weeks, Ca\,{\sc ii} NIR triplet absorption feature gets stronger. At around $+$23 day post-maximum, lines due to Cr\,{\sc ii} ($\sim$4600 \AA), Fe\,{\sc ii} ($\sim$5200 \AA), Co\,{\sc ii} ($\sim$5900 \AA, 6500 \AA), Fe\,{\sc ii} ($\sim$ 6100 \AA, 7000 \AA) can be clearly identified. In the post-maximum phase ($\sim +$10.5~d) the spectrum of SN~2020sck has more similarity with SN~2005hk. While SN~2019muj shows features due to Fe\,{\sc ii} and Co\,{\sc ii} beyond 6500 \AA, those are absent in SN~2020sck. This indicates that SN~2020sck has higher temperature than SN~2019muj in this phase.

The velocity evolution provides clues to the distribution of the elements in the ejecta and hence the explosion physics. The velocity of the spectral lines of SN~2020sck has been measured by fitting a Gaussian function to the absorption minimum of the corresponding lines. In the pre-maximum phase we fit Gaussian functions to Fe\,{\sc iii} ($\lambda$4420), Fe\,{\sc iii} ($\lambda$5156) and Si\,{\sc ii} ($\lambda$6355). We find the velocity of Si\,{\sc ii} to be 5712 $\pm$ 200 km~s$^{-1}$ and that of Fe\,{\sc iii} ($\lambda$4420) and Fe\,{\sc iii} ($\lambda$5156) to be 6610 $\pm$ 180~km s$^{-1}$ and 6649 $\pm$ 200~km s$^{-1}$ respectively. The Fe lines have velocities $\sim$ 800 km~s$^{-1}$ higher than those of Si\,{\sc ii}. For SN~2007qd \citep{2010ApJ...720..704M}, SN~2014ck \citep{2016MNRAS.459.1018T} the Fe lines are 800 km~s$^{-1}$ and 1,000 km~s$^{-1}$ higher than  Si\,{\sc ii} respectively. This trend has been also seen for other SNe Iax - like SN~2005hk \citep{2007PASP..119..360P} and SN~2010ae \citep{2014A&A...561A.146S}. This observation implies that fully burned materials are present in all the layers in the ejecta and that it supports an explosion mechanism that produces extensive mixing \citep{2007PASP..119..360P}. Around maximum, the velocity of Si\,{\sc ii}, C\,{\sc ii} ($\lambda$6580) and Ca\,{\sc ii} ($\lambda$3945) are 5185 km~s$^{-1}$, 5211 km~s$^{-1}$ and, 5308~km s$^{-1}$ respectively. However, the velocity of Fe\,{\sc iii} ($\lambda$5156) is 5558 $\pm$ 170 km~s$^{-1}$. The velocity of Fe\,{\sc iii} ($\lambda$4420) cannot be measured as it gets blended with other lines around the maximum. To understand the density profile and distribution of elements in the ejecta we compare the observed spectrum of SN~2020sck with synthetic spectrum generated using \texttt{SYN++} and \texttt{TARDIS}.

\section{Spectral Modeling} \label{Spectral Modeling}

The line velocities for SNe Iax are low and hence the spectral features post-maximum are easily identifiable than those for SNe Ia. The spectral features were identified using the parametrized spectrum synthesis code \texttt{SYN++} \citep{2011PASP..123..237T}. The code makes simple assumptions of homologous expansion of the ejecta in a spherically symmetric distribution. A synthetic spectrum is generated by assuming a well defined sharp photosphere that emits a continuous blackbody spectrum. Line formation occurs due to resonant scattering by assuming Sobolev approximation. The code can be used for line identifications, estimating the photospheric velocity and the velocity interval over which lines due to each ion are formed. The fit parameters are the temperature of the blackbody continuum ($T_{\rm BB}$), velocity of the photosphere ($v_{\rm phot}$), the minimum and maximum velocity of the line forming region ($v_{\rm min}$ \& $v_{\rm max}$), the optical depth of the ions ($\tau$), the Boltzmann excitation temperature ($T_{\rm exc}$) and the e-folding velocity ($aux$). A line is considered to be detached if the minimum velocity exceeds the photospheric velocity. The spectrum at $+$1.4 d has been compared with the synthetic spectrum to identify the lines and estimate the velocities. 

\begin{table*}
\centering
\renewcommand{\arraystretch}{1.1}
\setlength{\tabcolsep}{8pt}
\caption{\texttt{SYN++} fit to the pre-maximum spectrum of SN~2020sck.}
\label{tab:syn++}
\begin{tabular}{l c c c c c c c c c c c c c}
\hline \hline
\multicolumn{14}{c}{Phase$^*$: $+$ 1.4~d~~~~~~\(v_{\rm{phot}}\): 5000~km~s$^{-1}$~~~~~~\(T_{\rm{BB}}\): 10500~K} \\
\noalign{\smallskip} \hline
Parameters  &  C\,{\sc ii} & C\,{\sc iii} & O\,{\sc i}\(\rm{_{PV}}\) & O\,{\sc i}\(\rm{_{DF}}\) & Na\,{\sc i} & Si\,{\sc ii} & Si\,{\sc iii} & S\,{\sc ii} & Ca\,{\sc ii} & Sc\,{\sc ii} & Fe\,{\sc ii} & Fe\,{\sc iii} & Co\,{\sc ii} \\
\hline \noalign{\smallskip}
log~(tau) &  -1.5 & -1.4 & -1.0 & -1.2 & -1.8 & -1.1 & -1.1 & -1.2 & -0.1 & -1.5 & -0.9 & -0.9 & -0.9 \\
\(v_{\rm{min}}\) ($\times \rm 10^{3}$ km~$\rm s^{-1}$) & 5.0 & 5.0 & 5.0 & 11.0 & 5.0 & 5.0 & 5.0 & 5.0 & 5.0 & 5.0 & 5.0 & 5.0 & 5.0  \\
\(v_{\rm{max}}\) ($\times \rm 10^{3}$ km~$\rm s^{-1}$) & 7.0 & 8.0 & 15.0 & 15.0 & 10.0 & 12.0 & 7.0 & 7.0 & 8.0 & 7.0 & 7.0 & 8.0 & 7.0  \\
aux ($\times \rm 10^{3}$ km~$\rm s^{-1}$)  & 9.0 & 3.0 & 4.5 & 3.0 & 5.0 & 2.5 & 5.0 & 5.0 & 4.0 & 2.5 & 5.0 & 4.0 & 5.0  \\
\(T_{\rm{exc}}\) ($\times 10^{3}$ K) & 15 & 15 & 10 & 10 & 10 & 7 & 13 & 5 & 5 & 15 & 10 & 10 & 10  \\
\noalign{\smallskip} \hline
\multicolumn{14}{l}{$^*$\footnotesize{Time since $B$-band maximum (JD 2459098.84)}.} \\
\multicolumn{14}{l}{\footnotesize{$v_{\rm phot}$: The photospheric velocity (km~s$^{-1}$)}.} \\
\multicolumn{14}{l}{\footnotesize{$T_{\rm BB}$}: The blackbody continuum temperature (K).}
\end{tabular}
\end{table*}

\begin{figure}
\centering
\includegraphics[width=\columnwidth]{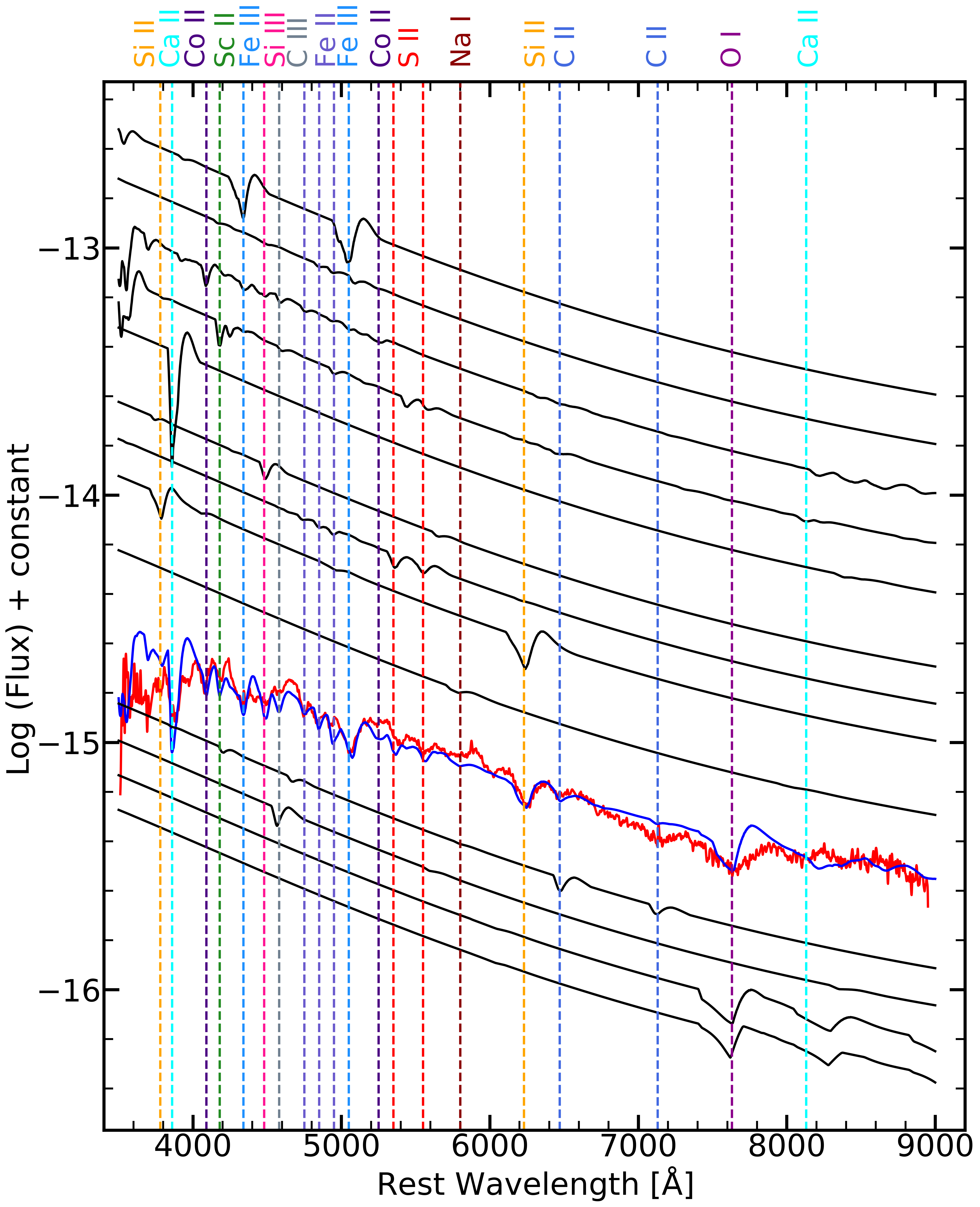}
\caption{Dereddened and redshift corrected spectra of SN~2020sck at $+$1.4 d since $B$-band maximum (shown in red). Overplotted in blue is the synthetic spectra calculated using \texttt{SYN++}. The contributions from each ion are shown by dashed vertical lines. The observed spectrum has been smoothed for visual clarity.}
\label{Fig12}
\end{figure}

The spectrum at $+$1.4~d was fit with a photospheric velocity ($v_{\rm phot}$) of 5,000 km s$^{-1}$ and a blackbody temperature ($T_{\rm BB}$) of 10,500 K. The spectrum has been fit with C\,{\sc ii}, C\,{\sc iii}, O\,{\sc i}, IME's like Na\,{\sc i}, Si\,{\sc ii}, Si\,{\sc iii}, S\,{\sc ii}, Ca\,{\sc ii} and IGE's like Sc\,{\sc ii}, Fe\,{\sc ii}, Fe\,{\sc iii}, Co\,{\sc ii}. To fit the broad O\,{\sc i} absorption feature around $\sim$7600 \AA\ , we used a photospheric component and a detached component at 11,000 km~s$^{-1}$. The velocity of the line forming region of C\,{\sc ii}, Si\,{\sc iii}, S\,{\sc ii}, Sc\,{\sc ii}, Fe\,{\sc ii}, Co\,{\sc ii} is between 5,000 km~s$^{-1}$ ($v_{\rm min}$) - 7,000 km~s$^{-1}$ ($v_{\rm max}$). The velocity of C\,{\sc iii} and Fe\,{\sc iii} is between 5,000 km~s$^{-1}$ - 8,000 km~s$^{-1}$. The Si\,{\sc ii} line has been fit with a velocity range of 5,000 km~s$^{-1}$ - 12,000 km~s$^{-1}$. This velocity range of IME's and IGE's show that the ejecta is mixed. To fit the C\,{\sc ii} and C\,{\sc iii} line profiles, an excitation temperature of 15,000 K has been used. The details of the fit are provided in Table~\ref{tab:syn++}. The detection of unburned carbon is of extreme importance as it can put constraint on the explosion mechanism as well as the progenitor system. The presence of C\,{\sc ii}, C\,{\sc iii} and O\,{\sc i} features hint towards thermonuclear explosion in a C-O white dwarf \citep{2010ApJ...708L..61F} in contrast to O-Ne-Mg white dwarf \citep{2013IAUS..281..253N}. The feature due to C\,{\sc iii} ($\lambda$4647) was also reported in SN~2014ck \citep{2016MNRAS.459.1018T}. Sc\,{\sc ii} feature was also identified in SN~2007qd \citep{2010ApJ...720..704M}, SN~2008ha \citep{2009AJ....138..376F}, SN~2010ae \citep{2014A&A...561A.146S}, SN~2014ck \citep{2016MNRAS.459.1018T}. All the major features are reproduced well in the synthetic spectrum. Fig.~\ref{Fig12} shows the \texttt{SYN++} fit to the $+$1.4~d spectrum of SN~2020sck.

\begin{figure}
\centering
\includegraphics[width=\columnwidth]{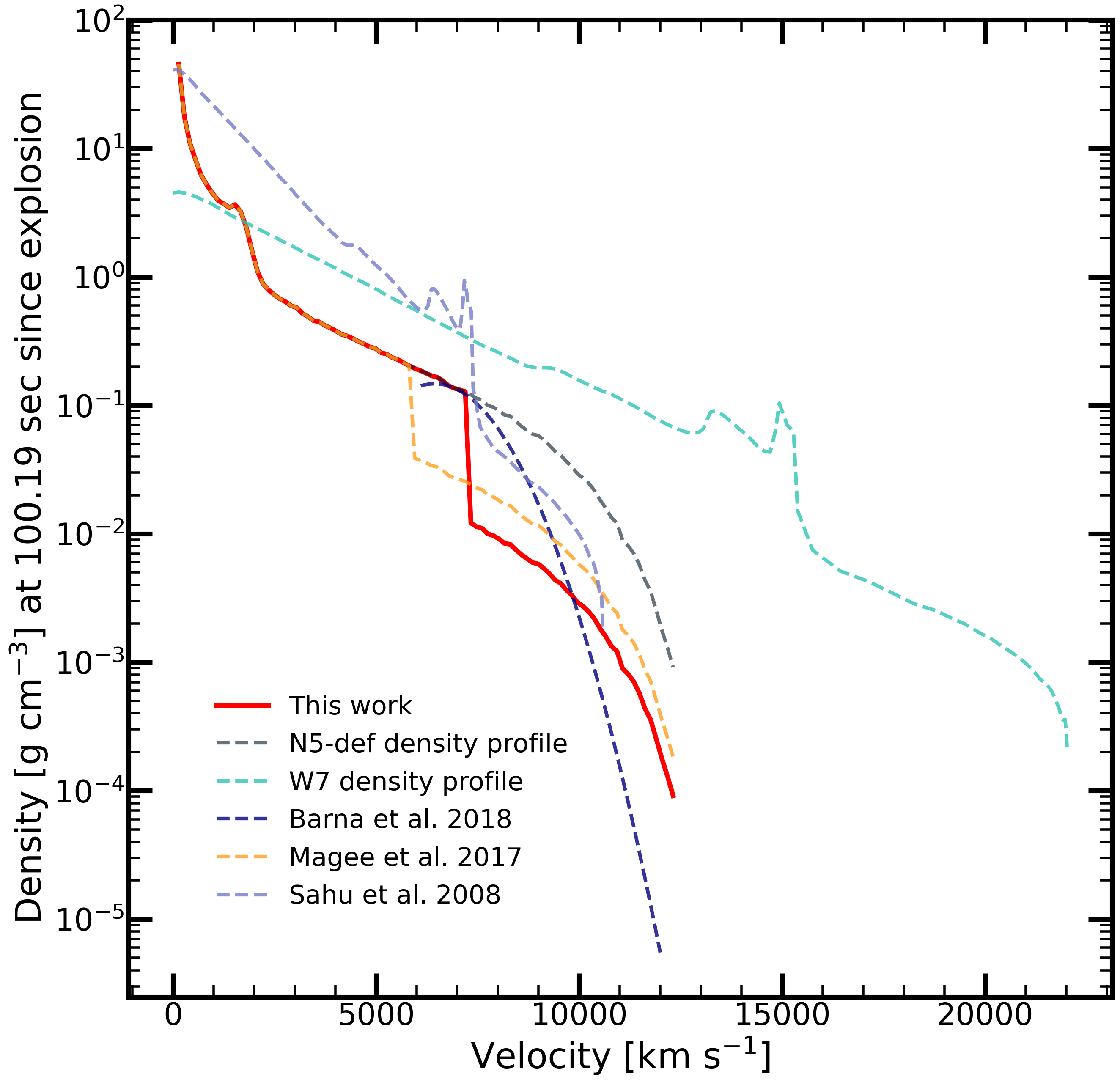}
\caption{Density against velocity plot (in red) used in \texttt{TARDIS} for generating the synthetic spectrum. The density is similar to N5-def density profile below 7200 km s$^{-1}$ and N5-def $\rho$ $\times$ 0.1 for velocity above 7200 km s$^{-1}$ . Also plotted for comparison - N5-def profile \citep{2014MNRAS.438.1762F}, W7 profile \citep{1984ApJ...286..644N}, density profile for SN~2005hk from \cite{2008ApJ...680..580S} and \cite{2018MNRAS.480.3609B}, density profile used for the study of PS1-12bwh \citep{2017A&A...601A..62M}.}
\label{Fig13}
\end{figure}

\begin{figure}
\centering
\includegraphics[width=\columnwidth]{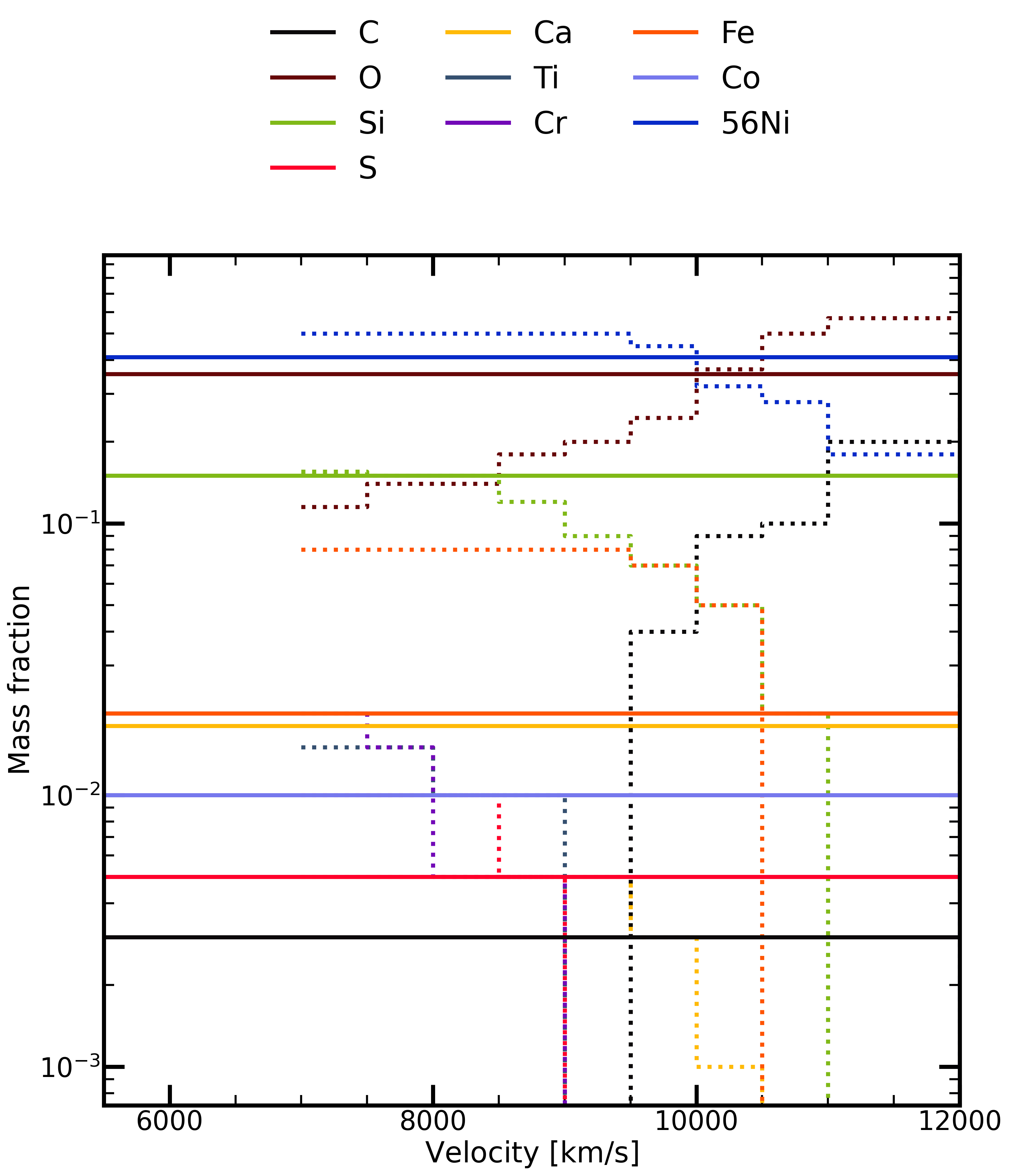}
\caption{Comparison of the uniform abundance of elements used for SN~2020sck with the stratified abundance for SN~2005hk \citep{2018MNRAS.480.3609B}. The solid lines show the mass fractions of the elements in the ejecta of SN~2020sck while the dotted lines in the same color shows the mass fraction in the ejecta of SN~2005hk.}
\label{Fig14}
\end{figure}

\begin{figure*}
\centering
\includegraphics[width=0.8\linewidth]{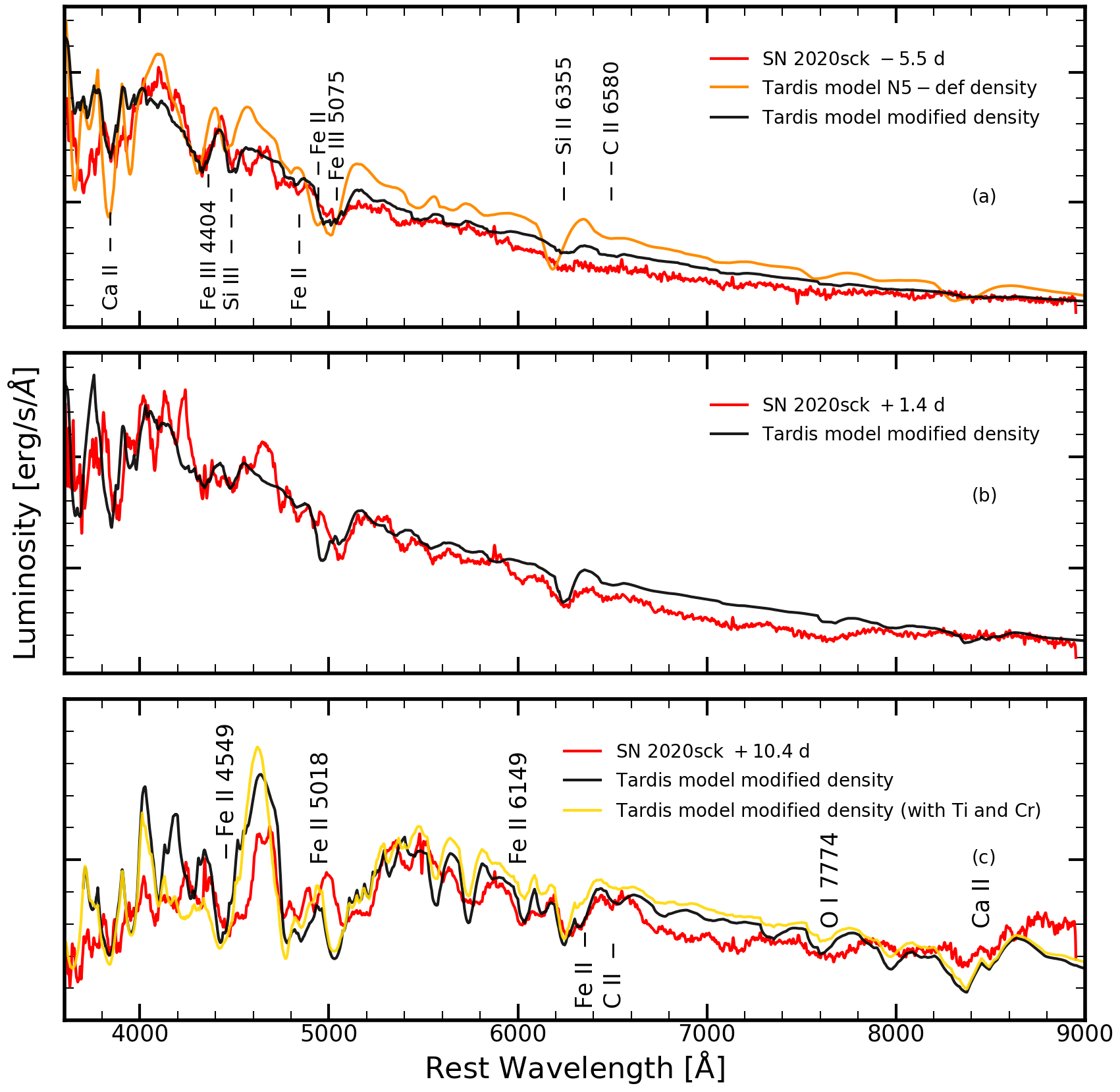}
\caption{Panel~(a) Spectrum of SN~2020sck at $-$5.5~d plotted along with synthetic spectrum generated using \texttt{TARDIS} with N5-def density profile (darkorange) and a modified N5-def density profile (black). Panel~(b) The $+$1.4~d spectrum of SN~2020sck plotted along with synthetic spectrum generated using the modified N5-def density profile. Panel~(c) Post-maximum $+$10.4~d spectrum of SN~2020sck compared with synthetic spectrum generated using N5-def modified density profile without (black) and with (gold) Ti and Cr. The angle-averaged density profile has been obtained from the Heidelberg Supernova Model Archive ({\sc hesma}). The observed spectra of SN~2020sck have been smoothed.}
\label{Fig15}
\end{figure*}

In order to put constraint on the explosion mechanism, perform line identification, estimate the abundance of the various elements ejected and get a knowledge of the ionisation state of the ejecta we compare the observed spectrum of SN~2020sck at $-$5.5~d, $+$1.4~d and $+$10.4~d since $B$-band maximum with synthetic spectrum generated using 1D Monte Carlo radiative transfer code \texttt{TARDIS} \citep{2014MNRAS.440..387K}. To generate a synthetic spectrum, \texttt{TARDIS} takes as input the luminosity of the SN ($L_{\rm SN}$ in log \(L_{\odot}\)), the time since explosion ($t_{\rm exp}$ in days), a density profile (density as a function of velocity), and uniform/stratified abundance. It assumes spherical symmetry, homologous expansion, a sharp well-defined photosphere and that the material in the computational domain defined by $v_{\rm inner}$ and $v_{\rm outer}$ is in radiative equilibrium. This makes the application of \texttt{TARDIS} limited to the photospheric phase. A synthetic spectrum is generated by considering a large number of Monte Carlo packets and tracing their propagation taking into account the interaction they make with the surrounding medium.

For generating the synthetic spectra, we considered the angle-averaged density profile of the three-dimensional pure deflagration explosion simulation (N5-def, \citealp{2014MNRAS.438.1762F}). In the N5-def explosion model, Fe and $^{56}$Ni are distributed to the outer parts of the ejecta. C and O are distributed in the entire ejecta and not limited to the outer regions. This indicates a mixed composition. We considered a uniform mass fraction of elements in the ejecta throughout the velocity interval.

\begin{table*}
\centering
\renewcommand{\arraystretch}{1.1}
\addtolength{\tabcolsep}{8.0pt}
\caption{Fit parameters of \texttt{TARDIS} model and comparison of ejecta composition with N5-def model \citep{2014MNRAS.438.1762F}.}
\label{tab:Tardis_fit}
\begin{tabular}{c c c c c c c c c c}
\hline \hline
X(C)  &  X(O)  &  X(Si)  &  X(S)  &  X(Ca)  &  X(Ti)  &  X(Cr)  &  X(Co)  &  X(Fe)  &  X(Ni) \\
\hline
\multicolumn{10}{c}{Phase$^*$: $-$ 5.5~d~~~~\(v_{\rm{inner}}\): 6800~km s$^{-1}$~~~~\(v_{\rm{outer}}\): 12000~km s$^{-1}$~~~~\(L_{\rm{SN}}\): 8.95~log \(L_{\rm{\odot}}\)~~~~\(t_{\rm{inner}}\): 11287~K} \\
\noalign{\smallskip} \hline
0.003 & 0.355 & 0.15 & 0.005 & 0.018 & 0.000 & 0.000 & 0.010 & 0.020 & 0.410 \\
\hline 
\multicolumn{10}{c}{Phase$^*$: $+$ 1.4~d~~~~\(v_{\rm{inner}}\): 6200~km s$^{-1}$~~~~\(v_{\rm{outer}}\): 12000~km s$^{-1}$~~~~\(L_{\rm{SN}}\): 9.05~log \(L_{\rm{\odot}}\)~~~~\(t_{\rm{inner}}\): 10033~K} \\
\noalign{\smallskip} \hline
0.003 & 0.355 & 0.15 & 0.005 & 0.018 & 0.000 & 0.000 & 0.010 & 0.020 & 0.410 \\
\hline
\multicolumn{10}{c}{Phase$^*$: $+$ 10.4~d~~~~\(v_{\rm{inner}}\): 5800~km s$^{-1}$~~~~\(v_{\rm{outer}}\): 12000~km s$^{-1}$~~~~\(L_{\rm{SN}}\): 8.85~log \(L_{\rm{\odot}}\)~~~~\(t_{\rm{inner}}\): 7780~K} \\
\noalign{\smallskip} \hline
0.003 & 0.200 & 0.080 & 0.005 & 0.018 & 0.020 & 0.020 & 0.010 & 0.180 & 0.410 \\
\hline
\multicolumn{10}{c}{N5-def model mean abundances} \\
\noalign{\smallskip} \hline
0.114 & 0.157 & 0.065 & 0.023 & 0.003 & 0.00 & 0.00 & 0.009 & 0.01 & 0.427 \\
\hline
\multicolumn{10}{l}{$^*$\footnotesize{Time since $B$-band maximum (JD 2459098.84)};~~~~\footnotesize{$v_{\rm inner}$: Inner velocity of the ejecta (km s$^{-1}$)}.} \\
\multicolumn{10}{l}{\footnotesize{$v_{\rm outer}$: Outer velocity of the ejecta (km s$^{-1}$)};~~~~~~~~\footnotesize{$L_{\rm SN}$: Luminosity of the SN (log \(L_{\rm{\odot}}\))}.} \\
\multicolumn{10}{l}{\footnotesize{$t_{\rm inner}$: Temperature of the photosphere (K)}.}
\end{tabular}
\end{table*} 

For comparing the observed spectrum at $-$5.5~d, we generate the synthetic spectrum with $t_{\rm exp}$=12.0 d, $L_{\rm SN}$=8.95 log \(L_{\odot}\) and a velocity interval of 8000 (\(v_{\rm{inner}}\))-12000 (\(v_{\rm{outer}}\)) km s$^{-1}$ (see panel (a) of Fig.\ref{Fig15}). However, we find that the absorption features are very strong. Also, the continuum seems to be bluer. This could be due to the higher density in the ejecta. Decreasing the \(v_{\rm{inner}}\) increases the optical depth further and increases the line strengths. We then fit the spectra, by considering a modified version of the N5-def density profile and a velocity interval of 6800 - 12,000 km~s$^{-1}$. In this case, the density profile in the outer ejecta ($v$ $>$ 7200 km s$^{-1}$) has been reduced (N5-def $\rho$ $\times$ 0.1). This steep change in the density profile has been supported by other studies (\citealt{2008ApJ...680..580S, 2017A&A...601A..62M, 2018MNRAS.480.3609B}). The innermost regions of the ejecta are denser as compared to the outermost region. In Fig.~\ref{Fig13} we compare the density profile used for SN~2020sck with the N5-def density profile \citep{2014MNRAS.438.1762F} and W7 profile \citep{1984ApJ...286..644N}. We also show the density profiles used for the study of SN~2005hk (\citealt{2008ApJ...680..580S, 2018MNRAS.480.3609B}), PS1-12bwh \citep{2017A&A...601A..62M}. In the case of SN~2005hk, \cite{2008ApJ...680..580S} homologously scaled the density profile to increase the density in the inner regions, while \cite{2018MNRAS.480.3609B} used an exponential density profile with a cut-off velocity $v_{\rm cut}$ chosen to match the deflagration density profiles. In PS1-12bwh, \cite{2017A&A...601A..62M} used N5-def density profile for velocity lower than 5800 km s$^{-1}$ and N5-def $\rho$ $\times$ 0.2 for velocities above 5800 km s$^{-1}$.

A uniform composition of elements throughout the entire ejecta ($v$ $\ge$ \(v_{\rm inner}\)) is supported by the mixed abundance structure in pure deflagration models. The \texttt{syn++} synthetic spectrum also indicates the elements are distributed throughtout the entire ejecta. In this case, we find the photospheric temperature to be $t_{\rm inner}$ = 11287 K, which is similar to that found by fitting a blackbody to the photometric spectral energy distribution (11037 K). The synthetic spectrum reproduces the more prominent lines due to Ca\,{\sc ii} (H $\&$ K), Fe\,{\sc iii} ($\lambda$4420),  Fe\,{\sc iii} ($\lambda$5156), S\,{\sc ii} and Si\,{\sc ii} ($\lambda$6355). C\,{\sc ii} ($\lambda$6578) feature is reproduced with a mass fraction X(C) = 0.003 while it is 0.114 in the N5-def model \citep{2014MNRAS.438.1762F}. 

To further investigate the effect of the density profile and the abundance structure, we compare the spectrum at $+$1.4~d with a synthetic spectrum generated with $t_{\rm exp}$=18.0 d, $L_{\rm SN}$=9.05 log \(L_{\odot}\) and a velocity interval of 6200 (\(v_{\rm{inner}}\)) - 12000 (\(v_{\rm{outer}}\)) km s$^{-1}$. We used the same mass fraction for the elements. The photospheric temperature (\(t_{\rm{inner}}\)) is 10033 K. This matches well with that found from the synthetic spectrum generated by \texttt{SYN++}, \(T_{\rm{phot}}\)= 10500 K. Here also, the absorption features due to C, Si, S, Fe and  Ca are reproduced well in the spectrum. However, the absorption feature around $\sim$4200 \AA\  due to Co are not reproduced (panel (b) in Fig.\ref{Fig15}).

The synthetic spectrum at $+$10.4~d has been generated with $t_{\rm exp}$=26.0 d, $L_{\rm SN}$=8.85 log \(L_{\odot}\) and a velocity interval of 5800 (\(v_{\rm{inner}}\)) - 12000 (\(v_{\rm{outer}}\)) km~s$^{-1}$. The photospheric temperature is 7,780 K. In this model, we consider two cases - (i) With Ti and Cr in the ejecta and (ii) Without Ti and Cr (panel (c) in Fig.\ref{Fig15}). Introducing Ti and Cr reduces the flux in the bluer region around $\sim$ 4300 \AA. In this phase we increase the mass fraction of Fe from X(Fe) = 0.02 to X(Fe) = 0.18. Similarly, we decrease the mass fraction of Si from X(Si) = 0.15 to X(Si) = 0.08. This means that the ejecta is entering into an Fe dominated phase. The absorption features due to  Fe\,{\sc ii} ($\lambda$4549),  Fe\,{\sc ii} ($\lambda$5018),  Fe\,{\sc ii} ($\lambda$6149),  Fe\,{\sc ii} ($\lambda$6247), Fe\,{\sc ii} ($\lambda$6456), C\,{\sc ii} ($\lambda$6578),  O\,{\sc i} ($\lambda$7774) and  Ca\,{\sc ii}-IR triplet are reproduced in the synthetic spectrum also. 

While three-dimensional deflagration models predict a mixed abundance structure, \cite{2018MNRAS.480.3609B} made a template based approach with stratified abundance structure to explore the ejecta of several bright SNe Iax. In the template model, the mass fraction of the IGE's and IME's decreases with velocity and C is tolerated only in the outermost regions. However, in this work we model the spectra using the same mass fraction over the velocity interval for the elements in the ejecta. This is in close resemblance to the three dimensional hydrodynamic simulations. Table~\ref{tab:Tardis_fit} lists the mass fractions of the elements used in the synthetic spectrum and comparison with the N5-def model mean abundances \citep{2014MNRAS.438.1762F}. In Fig.~\ref{Fig14} we compare the uniform abundance of the elements in the ejecta of SN~2020sck with the stratified abundance structure for SN~2005hk \citep{2018MNRAS.480.3609B}.

From the \texttt{TARDIS} models, we find that the density in the inner regions is higher than the outer regions. From the \texttt{syn++} synthetic spectrum we find that C, O and Fe group elements are located in the ejecta between 5000 km~s$^{-1}$ - 8000 km~s$^{-1}$. Using a uniform composition of the elements between 5800 km~s$^{-1}$ - 12000 km~s$^{-1}$ in the ejecta we confirm that most of the prominent features of C, O, Fe, Si and Ca can be reproduced in the \texttt{TARDIS} synthetic spectrum as well. However, some features due to C\,{\sc iii} ($\sim$ 4600 \AA), Co\,{\sc ii} ($\sim$ 4100 \AA), Fe\,{\sc ii} ($\sim$ 4800 \AA) are reproduced well in the \texttt{syn++} model but not in the \texttt{TARDIS} model. The analyses presented here indicate the elements in the ejecta of SN~2020sck are mostly mixed and support an explosion that is probably due to pure deflagration of a C-O white dwarf.

\section{Host Galaxy} \label{Host}

\begin{figure}
\centering
\includegraphics[width=\columnwidth]{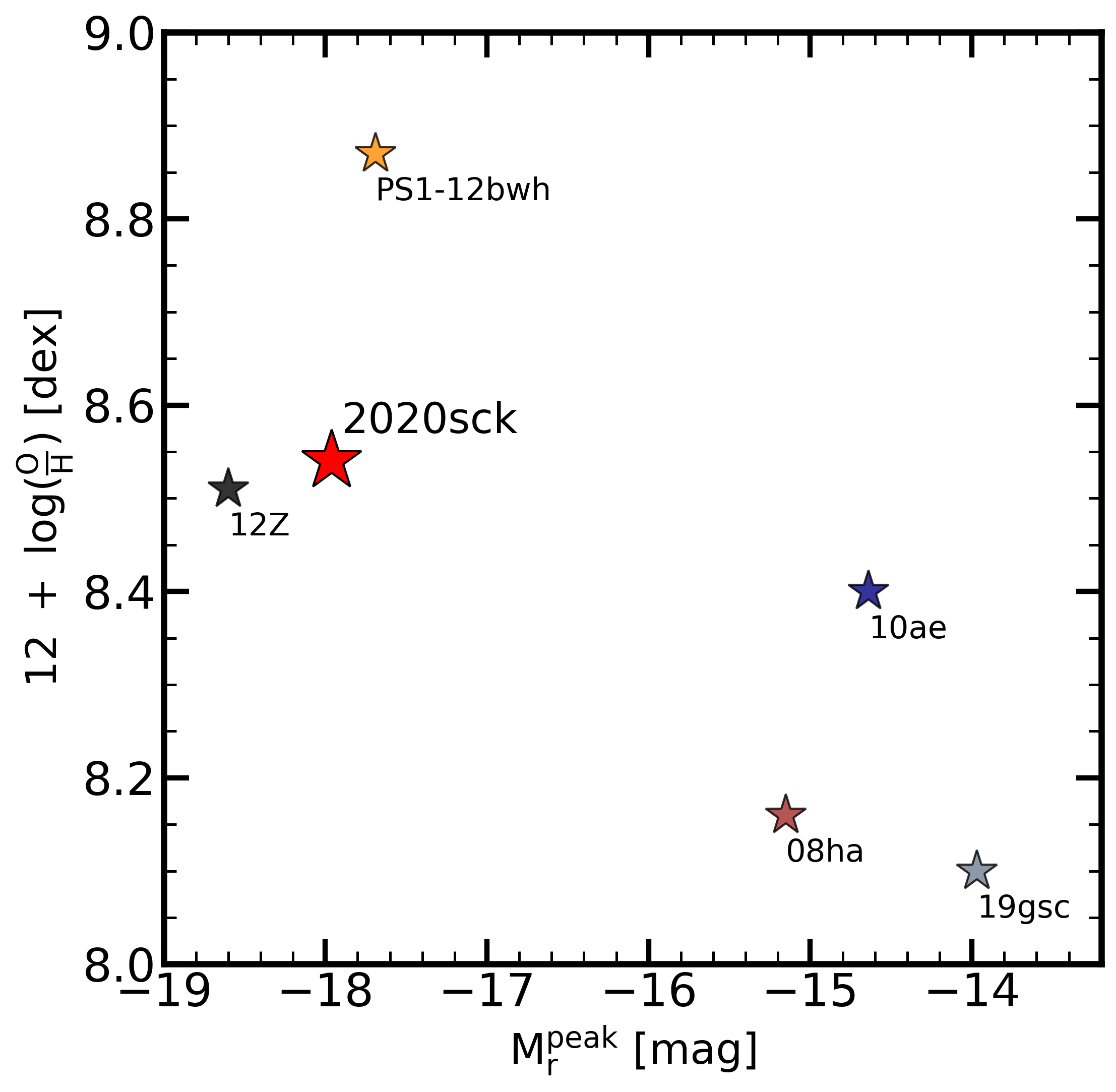}
\caption{$\rm 12 + log (O/H)$ plotted against peak magnitude in $r$-band for a sample of SNe Iax. The luminous objects have higher oxygen abundance.}
\label{Fig16}
\end{figure}

The metallicity of the host galaxy 2MASX J01103497+0206508 can be determined from the narrow emission features in the SN spectrum. We fit Gaussian profiles to the narrow H$\rm \alpha$ ($\lambda$6563) and [N\,{\sc ii}] $\lambda$6583. Using the empirical relation derived by \citep{2004MNRAS.348L..59P} with the N2 index \(\rm (log{\frac{[N\,{\sc II}]\ \lambda 6583}{H \rm \alpha\ \lambda 6563}})\), we find the oxygen abundance to be 12 + \(\rm log(O/H) = 8.54 \pm 0.05\ dex\). Metallicity values of a handful of SNe Iax have been obtained using the same relation of \cite{2004MNRAS.348L..59P} - SN~2008ha, SN~2010ae, SN~2012Z, PS1-12bwh, SN~2019gsc. These metallicity values for the comparison SNe are also listed in Table~\ref{tab:SNIax_sample}. SN~2008ha ($L_{\rm peak}$ = 9.5 $\times$ 10$^{40}$ erg s$^{-1}$) and SN~2019gsc ($L_{\rm peak}$ = 7.4 $\times$ 10$^{40}$ erg s$^{-1}$) have metal poor environments and low peak luminosity. There could be a relation between metallicity of the host galaxy and peak supernova magnitude, with low luminosity SNe Iax having lower metallicity (Fig.~\ref{Fig16}). However, there exist no clear correlation which demonstrates that SNe Iax tend to form in sub/super-solar metallicity environments \citep{2017A&A...601A..62M}. 

Using the spectrum at +23 d from the $B$-band maximum obtained with a $1.92^{\prime\prime}$ slit, we find the star formation (SFR) of the H\,{\sc ii} region near the SN. From the luminosity of the H$\alpha$ ($\lambda$6563) line, we derive an SFR \citep{1998ARA&A..36..189K} of 0.09 \(M_{\odot}\)~yr$^{-1}$. Young massive stars ($\ge$ 10 \(M_{\odot}\)) mostly contribute to the integrated line flux. From the forbidden [O\,{\sc ii}]($\lambda$3727) line luminosity, we derive a star formation rate of 0.02 \(M_{\odot}\)~yr$^{-1}$. The SFR derived from [O\,{\sc ii}] is less precise and suffers from systematic errors due to extinction \citep{1998ARA&A..36..189K}. In comparison, \cite{2009AJ....138..376F} derived a star formation rate of 0.07 \(M_{\odot}\)~yr$^{-1}$ for the host galaxy of SN~2008ha using far-infrared luminosity.

\section{Explosion Models} \label{Explosion Models}

\begin{table*}
\centering
\setlength{\tabcolsep}{7pt}
\caption{Fit paramaters of 1D radiation diffusion model (Eqn~\ref{eq:Arnett}) to the sample of SNe Iax.}
\label{tab:SNIax_Arnett}
\begin{tabular}{l c c c c c c c}
\hline
SN      &   \(M_{\rm{Ni}}\)       &   \(t_{\rm{lc}}\)  &    \(t_{\rm{\gamma}}\)       &  \(JD_{\rm{exp}}\)     & \(M_{\rm{ej}}\)  & \(v_{\rm{exp}}\) &  \(E_{\rm{K}}\)\\
    &   (M\(_{\odot}\))   &       (days)         &     (days) &         &      (M\(_{\odot}\))     &  (km~s\(^{-1}\))     &   (\(\times\) 10\(^{51}\) erg)       \\
\hline
SN~2002cx & \(0.07^{+0.01}_{-0.01}\) & \(12.29^{+2.12}_{-1.78}\) & \(42.6^{+3.1}_{-4.2}\) & \(2452402.43^{+1.12}_{-1.85}\) & \(0.69^{+0.20}_{-0.24}\) & 6000 & \(0.15^{+0.04}_{-0.05}\) \\
SN~2005hk & \(0.14^{+0.01}_{-0.01}\) & \(15.00^{+1.84}_{-1.58}\) &  \(52.8^{+4.3}_{-2.9}\) & \(2453669.65^{+1.02}_{-1.17}\) & \(1.13^{+0.27}_{-0.23}\) & 6500 & \(0.29^{+0.06}_{-0.06}\) \\
SN~2008ha & \(0.004^{+0.000}_{-0.000}\) & \(9.30^{+1.79}_{-0.52}\) & \(27.3^{+0.1}_{-0.2}\) & \(2454773.02^{+0.36}_{-2.03}\) & \(0.13^{+0.04}_{-0.01}\) & 2000 & \(0.003^{+0.0009}_{-0.0003}\) \\
SN~2010ae & \(0.006^{+0.001}_{-0.001}\) & \(5.43^{+1.79}_{-0.52}\) & \(38.5^{+2.9}_{-1.5}\) & \(2455246.48^{+0.5}_{-1.1}\) & \(0.13^{+0.08}_{-0.02}\) & 5500 & \(0.02^{+0.01}_{-0.00}\) \\
SN~2012Z & \(0.17^{+0.01}_{-0.01}\) & \(12.97^{+0.66}_{-0.50}\) & \(44.4^{+1.2}_{-1.2}\) & \(2455954.32^{+0.31}_{-0.37}\) & \(1.04^{+0.1}_{-0.09}\) & 8000 & \(0.39^{+0.08}_{-0.14}\) \\
SN~2014ck & \(0.03^{+0.01}_{-0.01}\) & \(8.32^{+0.58}_{-0.57}\) & \(39.6^{+1.9}_{-1.4}\) & \(2456835.55^{+0.25}_{-0.24}\) & \(0.16^{+0.02}_{-0.02}\) & 3000 & \(0.009^{+0.001}_{-0.001}\) \\
SN~2014ek & \(0.12^{+0.01}_{-0.01}\) & \(16.10^{+1.21}_{-2.41}\) & \(27.0^{+1.37}_{-0.37}\) & \(2456944.12^{+1.39}_{-0.63}\) & \(0.90^{+0.14}_{-0.27}\) & 4500 & \(0.11^{+0.02}_{-0.03}\) \\
SN~2019gsc & \(0.0019^{+0.000}_{-0.000}\) & \(9.85^{+0.25}_{-0.26}\) & \(42.9^{+0.37}_{-0.43}\) & \(2458627.24^{+0.3}_{-0.3}\) & \(0.26^{+0.098}_{-0.084}\) & 3500 & \(0.019^{+0.007}_{-0.006}\) \\
SN~2019muj & \(0.017^{+0.001}_{-0.001}\) & \(6.2^{+0.63}_{-0.93}\) & \(28.4^{+0.65}_{-0.70}\) & \(2458699.52^{+0.65}_{-0.46}\) & \(0.14^{+0.05}_{-0.03}\) & 5000 & \(0.024^{+0.006}_{-0.005}\) \\
\hline
\multicolumn{8}{l}{Note: All the parameters are explained in Section~\ref{Ni_mass}}
\end{tabular}
\newline \newline
\end{table*}

Considering the single degenerate \(M_{\rm ch}\) white dwarf explosion, we discuss three models based on the propagation of the burning front through the white dwarf to explain the explosion of SN 2020sck and similar SNe Iax.

\begin{figure}
\centering
\includegraphics[width=\columnwidth]{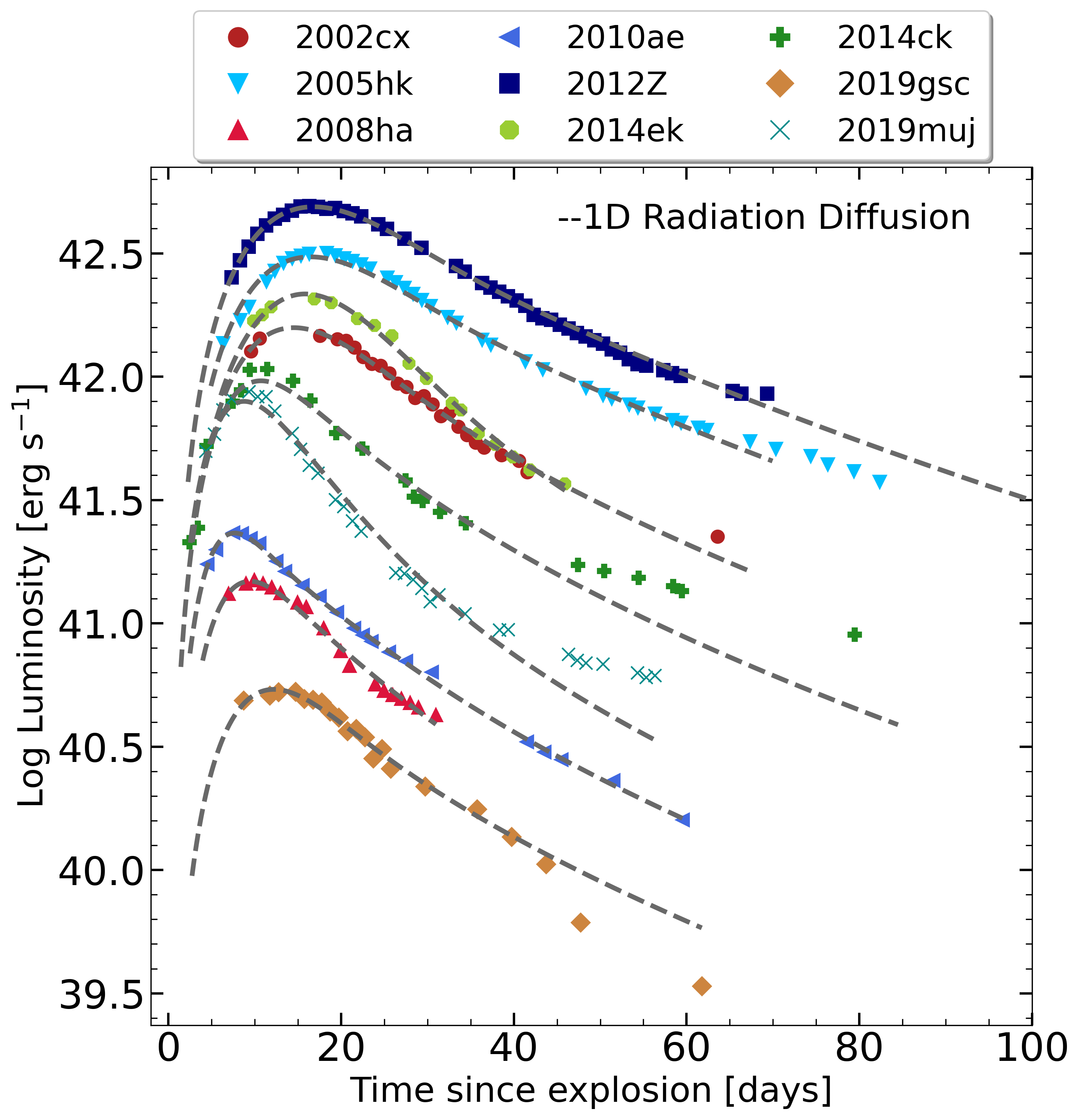}
\caption{1D radiation diffusion model fit to a sample of SNe Iax. A quasi-bolometric light curve has been constructed by integrating the flux from 3000 \AA\ to 9500 \AA\ for all objects. See Table~\ref{tab:SNIax_Arnett} for the fit parameters.}
\label{Fig17}
\end{figure}

First, we consider the deflagration-to-detonation (DDT) transition models. In these models the deflagration flame transitions into a detonation due to turbulent velocity fluctuations. In three dimensional simulations of \(M_{\rm{ch}}\) white dwarfs, a range of observed luminosity can be produced (\citealt{2013MNRAS.429.1156S, 2013MNRAS.436..333S}). By varying the deflagration strength and central density of the white dwarf a set of models have been generated which can account for the observed properties of SNe Ia. In this set of models, the explosions have been generated by considering a distribution of ignition points. The models with greater deflagration strengths produce a lesser amount of $^{56}$Ni because the white dwarf expands more before detonation sets in. The models produce a range of $^{56}$Ni mass of 0.32 - 1.1 \(M_{\odot}\). This mass range is higher than that found for SN~2020sck (0.13 \(M_{\odot}\)). We also considered a sample of SNe Iax and constructed the quasi-bolometric light curve (3000 \AA~to 9500 \AA). The quasi-bolometric light curves have been fit with the modified radiation diffusion model (Eq.~\ref{eq:Arnett}). Table~\ref{tab:SNIax_Arnett} shows the fit parameters of the radiation diffusion model to the SNe Iax sample considered here, and Fig.~\ref{Fig17} shows the fit of Eq.~\ref{eq:Arnett} to the quasi-bolometric light curves of the sample. The range of $^{56}$Ni (0.004 \(M_{\odot}\) - 0.17 \(M_{\odot}\)) estimated from the fitting is lower than that inferred from the DDT models.  The kinetic energy produced by the DDT models (\(E_{\rm{K}}\) = 1.20 - 1.67 $\times$ 10\(^{51}\) erg) is also higher than that observed for SNe Iax. Through our fit to the bolometric light curve, we find the range of kinetic energy between 0.003 $\times$ 10\(^{51}\) erg (SN~2008ha) - 0.39 $\times$ 10\(^{51}\) erg (SN~2012Z). The peak magnitude for the DDT model with the weakest deflagration N1 (with one ignition spot) is \(M_{\rm{B}}\) = -19.93. But, the model with the strongest deflagration N1600 (with 1600 deflagration spots and a central density $\rho_{\rm c}$ = 2.9 $\times$ 10$^{9}$ gm~$\rm cm^{-3}$) can produce the luminosity of \(M_{\rm B}\) = -18.26 mag observed in the brighter Iax. However, the IME production for model N1600 is too large (M(Si) = 0.36 \(M_{\odot}\)) and the velocity higher compared to SNe Iax. The $(B-V)$ color for the DDT models at $B$-maximum is too red (0.15~mag - 0.56~mag) compared to SN~2020sck $(B-V)$ = $-$0.08 mag. The DDT models do not seem to reproduce most of the observed properties of explosion for SN~2020sck and the sample SNe Iax.

Next, we consider the pulsational delayed detonation (PDD) model. Due to slow deflagration in a white dwarf, it expands but remains bound. As the burning stops, the infalling C-O layer compresses the IGE-rich mixed layers. As a result, detonation is triggered by compression and ignition (\citealt{1991A&A...245L..25K, 1995ApJ...444..831H}). In the one-dimensional case, several models have been generated by varying the transition density. This gives rise to a range of $^{56}$Ni mass (0.12 - 0.66 \(M_{\odot}\)). The $^{56}$Ni mass found for SN~2020sck matches with the model PDD5 (the transition density for this model at which the deflagration is turned to detonation is $\rho_{\rm tr}$ = 0.76 $\times$ 10$^{7}$ gm~$\rm cm^{-3}$) for which the amount of $^{56}$Ni produced is 0.12 \(M_{\odot}\). However, the average expansion velocity for this model is 8,400 km~s$^{-1}$. This is higher than that found in SN~2020sck ($\sim$5,000 km~s$^{-1}$). The $(B-V)$ color in the PDD5 model is 0.44 mag which is redder than that for SN~2020sck ($-$0.08 mag).

In the PDD models, the kinetic energy varies from 0.34 - 1.52 $\times$ 10\(^{51}\) erg. The range of velocity and hence, kinetic energy is also observed in SNe Iax. The extreme case PDD535 ($\rho_{\rm tr}$ = 0.45 $\times$ 10$^{7}$ gm~$\rm cm^{-3}$) \citep{1995ApJ...444..831H} has low $^{56}$Ni mass (0.16 \(M_{\odot}\)) and low average expansion velocities $\sim$ 4,500 km~s\(^{-1}\). Due to the pulsation, the material which is falling back interacts with the outgoing detonation wavefront. As a result, a dense shell of mass is formed surrounded by fast-moving layers. These fast-moving layers take away some kinetic energy and decelerate the inner parts of the expanding ejecta. This result in lower expansion velocities. However, in the case of PDD535, the Fe and Ni layers are located below 4,000 km~s\(^{-1}\). By comparing the synthetic spectra with the observations of SN~2020sck, we find that the Fe and Ni line forming layers are present in the outer parts of the ejecta ($\ge$ 7,000 km~s$^{-1}$). The $(B-V)$ color is 0.60 mag which is redder than that for SN~2020sck. Hence, the PDD models also do not reproduce some of the observed properties of SN~2020sck.

Previous study by \cite{2014MNRAS.438.1762F} have shown that most of the observed properties of SNe Iax class (brighter and intermediate luminosity) can be successfully described by pure deflagrations of \(M_{\rm{ch}}\) C-O white dwarf. \cite{2014MNRAS.438.1762F} have generated a set of models by varying the deflagration strength (changing the number of ignition spots). These models produce a range of $^{56}$Ni mass (0.03 - 0.34 \(M_{\odot}\)), with the peak $B$-band magnitude varying from -16.55 (N1) to -18.11 (N1600). The models also produce mixed abundance distribution in which Fe and Ni can be present in the outer layers of the ejecta. Models with weak and intermediate deflagration strengths (N1 - N100) produce lesser ejecta and a bound remnant. Comparing with the various models, we find that the explosion properties of the N5-def model with five ignition points (with the central density of the white dwarf being $\rho_{\rm c}$ = 2.9 $\times$ 10$^{9}$ gm~$\rm cm^{-3}$) matches well with that found by fitting the bolometric light curve of SN~2020sck with the radiation diffusion model. Also, the N5-def density profile with a steep decrease in the outer layers can successfully reproduce the observed spectral features. The rise-times to $B$-maximum for the pure deflagration models are less (7.6~d - 14.4~d) compared to SN~2020sck, but fit the range observed in the sample SNe Iax. 

In Table~\ref{tab:explosion}, we compare various explosion scenarios of single degenerate \(M_{\rm ch}\) white dwarfs and show which model can best explain the observed properties of SN~2020sck and the sample SNe Iax. Based on this, like in the previous studies by \cite{2003PASP..115..453L}, \cite{2007PASP..119..360P}, \cite{2014MNRAS.438.1762F} we conclude that the pure deflagration models explain most of the observed paramaters of SNe Iax.

\begin{table*}
\centering
\setlength{\tabcolsep}{10pt}
\caption{Comparing the explosion properties of SN~2020sck with single degenerate \(M_{\rm ch}\) white dwarf explosion models.}
\label{tab:explosion}
\begin{tabular}{l c c c c c c c}
\hline \hline
Explosion Model   & Velocity &  Peak Magnitude & Color &  Rise-Time & Spectra & $^{56}$Ni & Ref.\\
\hline
Deflagration-to-Detonation & \xmark  & \xmark & \xmark & \cmark & \xmark & \xmark & 1\\
Pulsational Delayed Detonation & \xmark & \xmark & \xmark & \cmark & \xmark & \xmark & 2\\
(Except PDD5 and PDD535) & & & & & & & \\
PDD5 & \xmark & \cmark & \xmark & \xmark & \xmark & \cmark & 2\\
PDD535 & \xmark & \cmark & \xmark & \xmark & \xmark & \cmark & 2\\
Pure Deflagration (N5-def) & \cmark & \cmark & \cmark & \xmark & \cmark & \cmark & 3\\
\hline \noalign{\smallskip}    
\end{tabular}
\newline
References:\
(1) \cite{2013MNRAS.429.1156S}, \cite{2013MNRAS.436..333S}; (2) \cite{1995ApJ...444..831H}; (3) \cite{2014MNRAS.438.1762F} \\
Note: If the model matches the observed property of SN~2020sck, we put a \cmark else \xmark
\end{table*}

\section{SUMMARY} \label{Summary}

In this work we establish that SN~2020sck is a supernova of type Iax with a \(\Delta m_{B}(15) = 2.03\pm0.05\) mag and \(M_{B}=-17.81\pm0.22\) mag. From the pre-maximum observations in the ZTF$-g$ band, we constrained the date of explosion as 2020 August 20 (JD = 2459082.4). The light curves in $R$ and $I$-bands do not show any secondary maximum. The $(B-V)$ color at maximum is $-$0.08 mag which is bluer compared to the sample of SNe Iax. By fitting the quasi-bolometric light curve as well as the blackbody corrected bolometric light curve of SN~2020sck with 1D radiation diffusion model we find 0.13 \(M_\odot\) and 0.17 \(M_\odot\) of $^{56}$Ni respectively. By comparing the quasi-bolometric light curve with angle-averaged bolometric light curve from three-dimensional pure deflagration models of C-O white dwarfs with varying deflagration strengths, we find similarity of SN~2020sck with N5-def model \citep{2014MNRAS.438.1762F}. The mass ejected in the explosion is 0.34 \(M_{\odot}\) with a kinetic energy of 0.05 $\times \rm 10^{51}$ erg.

The spectral characteristics of SN~2020sck are similar to SN~2005hk and SN~2019muj. The comparison of the near-maximum spectrum of SN~2020sck with \texttt{SYN++} shows the presence of higher ionization states of elements like C\,{\sc iii}, Si\,{\sc iii}, Fe\,{\sc iii} indicating a hot photosphere. The presence of unburned C and O points towards a C-O white dwarf progenitor. Fe lines are found at higher velocities than IME's indicating that the ejecta is mixed. Angle-averaged one-dimensional density profile of pure deflagration explosion of \(M_{\rm ch}\) white dwarf with a steep decrease in the outer layers of the ejecta can successfully reproduce the prominent absorption features in the spectra of SN~2020sck. The metallicity of the host galaxy of SN~2020sck is similar to SN~2012Z \citep{2015ApJ...806..191Y} which exploded in a spiral galaxy. More studies of SNe Iax will help to understand the correlation with their host galaxy environment.

\section*{Acknowledgements}

We thank the anonymous referee for carefully going through the manuscript and providing detailed comments that helped in improving the content of the paper. We thank the staff of IAO, Hanle and CREST, Hosakote that made the observations possible. This work made use of data from the GROWTH-India Telescope (GIT) set up by the Indian Institute of Astrophysics (IIA) and the Indian Institute of Technology Bombay (IITB) with funding from DST-SERB and IUSSTF. It is located at IAO. We acknowledge funding by the IITB alumni batch of 1994, which partially supports operations of the telescope. The facilities at IAO and CREST are operated by the Indian Institute of Astrophysics, Bengaluru, an autonomous Institute under Department of Science and Technology, Government of India. We also thank the observers of HCT who shared their valuable time for Target of Opportunity (ToO) observations during the initial follow up. HK thanks the LSSTC Data Science Fellowship Program, which is funded by LSSTC, NSF Cybertraining Grant \#1829740, the Brinson Foundation, and the Moore Foundation; his participation in the program has benefited this work. Nayana A.J. would like to acknowledge DST-INSPIRE Faculty Fellowship (IFA20-PH-259) for supporting this research. This work is also based on observations obtained at the 3.6m Devasthal Optical Telescope (DOT), which is a National Facility run and managed by Aryabhatta Research Institute of Observational Sciences (ARIES), an autonomous Institute under Department of Science and Technology, Government of India. This work has made use of the NASA Astrophysics Data System\footnote{\url{https://ui.adsabs.harvard.edu/}} ({\sc ADS}), the NASA/IPAC extragalactic database\footnote{\url{https://ned.ipac.caltech.edu/}} ({\sc NED}) and NASA/IPAC Infrared Science Archive ({\sc IRSA})\footnote{\url{https://irsa.ipac.caltech.edu/applications/DUST/}} which is operated by the Jet Propulsion Laboratory, California Institute of Technology. We acknowledge, Weizmann Interactive Supernova Data REPository\footnote{\url{https://wiserep.weizmann.ac.il/}} (WISeREP), \citep{2012PASP..124..668Y}. This research made use of \texttt{TARDIS}, a community-developed software package for spectral synthesis in supernovae (\citealt{2014MNRAS.440..387K, 2019zndo...2590539K}). The development of \texttt{TARDIS} received support from the Google Summer of Code initiative and from ESA's Summer of Code in Space program. \texttt{TARDIS} makes extensive use of \texttt{Astropy} and \texttt{PyNE}. This work made use of the Heidelberg Supernova Model Archive ({\sc hesma})\footnote{\url{https://hesma.h-its.org}}. The analysis has made use of the following software and packages - (i) {\small \it{Image Reduction and Analysis Facility}} (\texttt{IRAF}), \cite{1993ASPC...52..173T}. (ii) \texttt{PyRAF}, \cite{2012ascl.soft07011S}. (iii) \texttt{NumPy}, \cite{van2011numpy}, (iv) \texttt{Matplotlib}, \cite{Hunter:2007}, (v) \texttt{Scipy}, \cite{2020SciPy-NMeth}, (vi) \texttt{pandas}, \cite{reback2020pandas}, (vii) \texttt{Astropy}, \cite{2013A&A...558A..33A}, (viii) \texttt{emcee}, \cite{2013ascl.soft03002F}, (ix) \texttt{Corner}, \cite{corner} (x) \texttt{SYN++}, \cite{2011PASP..123..237T}, (xi) \texttt{TARDIS}, \cite{2014MNRAS.440..387K}. 

\section*{Data Availability}

The observed data (reduced) presented in this work and also the results of the analyses obtained based on open-source resources like \texttt{TARDIS}, \texttt{syn++} is available online at Zenodo (doi:\url{https://doi.org/10.5281/zenodo.5619721}). The reduced spectra will also be made available in the WISeREP archive \citep{2012PASP..124..668Y}. Raw data (observed) can be made available by the first author on reasonable request.

\bibliography{Bibliography}
\bibliographystyle{aasjournal}

\newpage
\appendix
\section{Tables}
\newpage

\begin{table*}
\centering
\renewcommand{\arraystretch}{1.1}
\caption{$UBVRI$ magnitudes of local standards in the field of SN 2020sck.}
\label{tab:standardlog}
\begin{tabular}{c c c c c c}
\hline \hline
      ID         &         $U$               &    $B$                &    $V$                &    $R$                &    $I$       \\
 \hline \noalign{\smallskip}
 1      &       16.156$\,\pm\,$0.061   &   16.047$\,\pm\,$0.010  &   15.374$\,\pm\,$0.004  &   15.013$\,\pm\,$0.009  &   14.592$\,\pm\,$0.012  \\
 2      &       17.069$\,\pm\,$0.061   &   17.019$\,\pm\,$0.010  &   16.442$\,\pm\,$0.004  &   16.139$\,\pm\,$0.009  &   15.759$\,\pm\,$0.012  \\
 3      &       18.229$\,\pm\,$0.063   &   17.016$\,\pm\,$0.010 &   15.544$\,\pm\,$0.004  &   14.521$\,\pm\,$0.009  &   13.345$\,\pm\,$0.012  \\
 4      &       15.991$\,\pm\,$0.061   &   15.199$\,\pm\,$0.009 &   14.177$\,\pm\,$0.004  &   13.639$\,\pm\,$0.009  &   13.059$\,\pm\,$0.012  \\
 5      &       18.522$\,\pm\,$0.065   &   17.763$\,\pm\,$0.011  &   16.809$\,\pm\,$0.004  &   16.254$\,\pm\,$0.009  &   15.685$\,\pm\,$0.014  \\
 6      &       14.292$\,\pm\,$0.061   &   14.235$\,\pm\,$0.009  &   13.671$\,\pm\,$0.004  &   13.382$\,\pm\,$0.009 &   13.027$\,\pm\,$0.011  \\
 7      &       15.054$\,\pm\,$0.061   &   15.096$\,\pm\,$0.009  &   14.513$\,\pm\,$0.004  &   14.208$\,\pm\,$0.010  &   13.816$\,\pm\,$0.012  \\
 8      &       19.227$\,\pm\,$0.069   &   17.865$\,\pm\,$0.011  &   16.452$\,\pm\,$0.004  &   15.560$\,\pm\,$0.010  &   14.655$\,\pm\,$0.012 \\
\noalign{\smallskip} \hline
\end{tabular}
\end{table*}
\newpage
\afterpage{
\begin{longtable}[hbt!]{ccccccc}
\caption{$UBVRI$ and \(r^{\prime}\) magnitudes of SN~2020sck from HCT and GIT.} \label{tab:photlog} \\

\hline \multicolumn{1}{c}{JD} & \multicolumn{1}{c}{Date} & \multicolumn{1}{c}{Phase$^*$} & \multicolumn{1}{c}{Filter} & \multicolumn{1}{c}{Magnitude (mag)} & \multicolumn{1}{c}{Error (mag)} \\ \hline 
\endfirsthead

\multicolumn{6}{c}%
{{\bfseries \tablename\ \thetable{} -- continued from previous page}} \\
\hline \multicolumn{1}{c}{JD} & \multicolumn{1}{c}{Date} & \multicolumn{1}{c}{Phase$^*$} & \multicolumn{1}{c}{Filter} & \multicolumn{1}{c}{Magnitude (mag)} & \multicolumn{1}{c}{Error (mag)} \\ \hline 
\endhead

\hline \multicolumn{6}{|r|}{{Continued on next page}} \\ \hline
\endfoot

\hline \hline
\endlastfoot
2459093.38 & 2020-08-31 & -5.37 &      $B$ &          16.677 &       0.013 \\
2459094.23 & 2020-09-01 & -4.53 &      $B$ &          16.656 &       0.059 \\
2459095.31 & 2020-09-02 & -3.47 &      $B$ &          16.489 &       0.035 \\
2459100.34 & 2020-09-07 &  1.47 &      $B$ &          16.379 &       0.029 \\
2459101.42 & 2020-09-08 &  2.54 &      $B$ &          16.429 &       0.021 \\
2459102.44 & 2020-09-09 &  3.54 &      $B$ &          16.502 &       0.021 \\
2459104.42 & 2020-09-11 &  5.48 &      $B$ &          16.683 &       0.018 \\
2459106.46 & 2020-09-13 &  7.49 &      $B$ &          16.985 &       0.019 \\
2459110.43 & 2020-09-17 & 11.40 &      $B$ &          17.762 &       0.019 \\
2459113.21 & 2020-09-20 & 14.13 &      $B$ &          18.212 &       0.013 \\
2459114.41 & 2020-09-21 & 15.31 &      $B$ &          18.373 &       0.012 \\
2459117.24 & 2020-09-24 & 18.09 &      $B$ &          18.720 &       0.013 \\
2459120.34 & 2020-09-27 & 21.13 &      $B$ &          18.945 &       0.024 \\
2459122.35 & 2020-09-29 & 23.11 &      $B$ &          19.240 &       0.033 \\
2459148.22 & 2020-10-15 & 48.56 &      $B$ &          19.862 &       0.034 \\
2459167.10 & 2020-11-13 & 67.16 &      $B$ &          20.031 &       0.021 \\
2459093.37 & 2020-08-31 & -5.38 &      $I$ &          16.847 &       0.009 \\
2459094.22 & 2020-09-01 & -4.54 &      $I$ &          16.796 &       0.014 \\
2459095.30 & 2020-09-02 & -3.48 &      $I$ &          16.638 &       0.013 \\
2459100.33 & 2020-09-07 &  1.47 &      $I$ &          16.415 &       0.016 \\
2459101.42 & 2020-09-08 &  2.54 &      $I$ &          16.375 &       0.009 \\
2459102.43 & 2020-09-09 &  3.53 &      $I$ &          16.389 &       0.016 \\
2459104.41 & 2020-09-11 &  5.48 &      $I$ &          16.383 &       0.019 \\
2459106.45 & 2020-09-13 &  7.48 &      $I$ &          16.404 &       0.011 \\
2459110.43 & 2020-09-17 & 11.39 &      $I$ &          16.414 &       0.009 \\
2459113.20 & 2020-09-20 & 14.12 &      $I$ &          16.447 &       0.006 \\
2459114.22 & 2020-09-21 & 15.12 &      $I$ &          16.498 &       0.013 \\
2459117.24 & 2020-09-24 & 18.09 &      $I$ &          16.591 &       0.011 \\
2459120.33 & 2020-09-27 & 21.13 &      $I$ &          16.760 &       0.008 \\
2459122.32 & 2020-09-29 & 23.08 &      $I$ &          16.873 &       0.012 \\
2459126.28 & 2020-10-03 & 26.98 &      $I$ &          17.082 &       0.029 \\
2459138.44 & 2020-10-15 & 38.94 &      $I$ &          17.646 &       0.025 \\
2459148.22 & 2020-10-25 & 48.55 &      $I$ &          17.973 &       0.012 \\
2459167.09 & 2020-11-13 & 67.11 &      $I$ &          18.410 &       0.015 \\
2459093.37 & 2020-08-31 & -5.38 &      $R$ &          16.800 &       0.007 \\
2459094.22 & 2020-09-01 & -4.55 &      $R$ &          16.711 &       0.010 \\
2459095.30 & 2020-09-02 & -3.48 &      $R$ &          16.592 &       0.014 \\
2459100.33 & 2020-09-07 &  1.46 &      $R$ &          16.378 &       0.013 \\
2459101.42 & 2020-09-08 &  2.53 &      $R$ &          16.364 &       0.012 \\
2459102.43 & 2020-09-09 &  3.53 &      $R$ &          16.410 &       0.007 \\
2459104.41 & 2020-09-11 &  5.48 &      $R$ &          16.426 &       0.009 \\
2459106.45 & 2020-09-13 &  7.48 &      $R$ &          16.486 &       0.011 \\
2459110.43 & 2020-09-17 & 11.39 &      $R$ &          16.605 &       0.008 \\
2459113.20 & 2020-09-20 & 14.12 &      $R$ &          16.693 &       0.007 \\
2459114.22 & 2020-09-21 & 15.12 &      $R$ &          16.756 &       0.013 \\
2459117.24 & 2020-09-24 & 18.09 &      $R$ &          16.884 &       0.012 \\
2459120.33 & 2020-09-27 & 21.13 &      $R$ &          17.091 &       0.008 \\
2459122.31 & 2020-09-29 & 23.08 &      $R$ &          17.228 &       0.008 \\
2459138.44 & 2020-10-15 & 38.94 &      $R$ &          18.022 &       0.019 \\
2459148.21 & 2020-10-25 & 48.55 &      $R$ &          18.225 &       0.013 \\
2459167.09 & 2020-11-13 & 67.11 &      $R$ &          18.688 &       0.014 \\
2459093.39 & 2020-08-31 & -5.36 &      $U$ &          16.214 &       0.103 \\
2459094.24 & 2020-09-01 & -4.52 &      $U$ &          16.062 &       0.279 \\
2459095.31 & 2020-09-02 & -3.47 &      $U$ &          16.010 &       0.243 \\
2459100.34 & 2020-09-07 &  1.47 &      $U$ &          16.145 &       0.086 \\
2459101.43 & 2020-09-08 &  2.54 &      $U$ &          16.213 &       0.053 \\
2459102.44 & 2020-09-09 &  3.54 &      $U$ &          16.348 &       0.068 \\
2459104.42 & 2020-09-11 &  5.49 &      $U$ &          16.509 &       0.144 \\
2459106.46 & 2020-09-13 &  7.49 &      $U$ &          16.918 &       0.100 \\
2459110.44 & 2020-09-17 & 11.40 &      $U$ &          17.845 &       0.054 \\
2459113.21 & 2020-09-20 & 14.13 &      $U$ &          18.540 &       0.026 \\
2459114.42 & 2020-09-21 & 15.32 &      $U$ &          18.636 &       0.044 \\
2459117.25 & 2020-09-24 & 18.10 &      $U$ &          19.188 &       0.060 \\
2459120.34 & 2020-09-27 & 21.14 &      $U$ &          19.457 &       0.069 \\
2459122.36 & 2020-09-29 & 23.12 &      $U$ &          19.176 &       0.227 \\
2459148.23 & 2020-10-25 & 48.56 &      $U$ &          20.300 &       0.123 \\
2459167.10 & 2020-11-13 & 67.12 &      $U$ &          21.365 &       0.157 \\
2459093.38 & 2020-08-31 & -5.37 &      $V$ &          16.835 &       0.009 \\
2459094.22 & 2020-09-01 & -4.54 &      $V$ &          16.755 &       0.012 \\
2459095.30 & 2020-09-02 & -3.48 &      $V$ &          16.634 &       0.022 \\
2459100.33 & 2020-09-07 &  1.47 &      $V$ &          16.394 &       0.015 \\
2459101.42 & 2020-09-08 &  2.54 &      $V$ &          16.395 &       0.009 \\
2459102.43 & 2020-09-09 &  3.53 &      $V$ &          16.404 &       0.019 \\
2459104.42 & 2020-09-11 &  5.48 &      $V$ &          16.465 &       0.016 \\
2459106.46 & 2020-09-13 &  7.49 &      $V$ &          16.541 &       0.022 \\
2459110.43 & 2020-09-17 & 11.40 &      $V$ &          16.839 &       0.014 \\
2459113.21 & 2020-09-20 & 14.13 &      $V$ &          17.056 &       0.008 \\
2459114.22 & 2020-09-21 & 15.13 &      $V$ &          17.168 &       0.005 \\
2459117.24 & 2020-09-24 & 18.09 &      $V$ &          17.389 &       0.010 \\
2459120.33 & 2020-09-27 & 21.13 &      $V$ &          17.609 &       0.012 \\
2459122.32 & 2020-09-29 & 23.09 &      $V$ &          17.741 &       0.017 \\
2459138.45 & 2020-10-15 & 38.95 &      $V$ &          18.786 &       0.015 \\
2459148.22 & 2020-10-25 & 48.55 &      $V$ &          18.596 &       0.014 \\
2459167.09 & 2020-11-13 & 67.11 &      $V$ &          18.917 &       0.010 \\ 
2459100.43 & 2020-09-07 &  1.56 &      \(r^{\prime}\) &          16.391 &       0.035 \\
2459101.42 & 2020-09-08 &  2.53 &      \(r^{\prime}\) &          16.408 &       0.012 \\
2459102.40 & 2020-09-09 &  3.50 &      \(r^{\prime}\) &          16.373 &       0.017 \\
2459103.41 & 2020-09-10 &  4.49 &      \(r^{\prime}\) &          16.385 &       0.016 \\
2459104.38 & 2020-09-11 &  5.45 &      \(r^{\prime}\) &          16.428 &       0.020 \\
2459105.39 & 2020-09-12 &  6.44 &      \(r^{\prime}\) &          16.436 &       0.020 \\
2459106.41 & 2020-09-13 &  7.44 &      \(r^{\prime}\) &          16.481 &       0.015 \\
2459108.39 & 2020-09-15 &  9.39 &      \(r^{\prime}\) &          16.520 &       0.014 \\
2459109.40 & 2020-09-16 & 10.38 &      \(r^{\prime}\) &          16.580 &       0.028 \\
2459110.36 & 2020-09-17 & 11.33 &      \(r^{\prime}\) &          16.589 &       0.034 \\
2459111.38 & 2020-09-18 & 12.33 &      \(r^{\prime}\) &          16.650 &       0.016 \\
2459112.39 & 2020-09-19 & 13.32 &      \(r^{\prime}\) &          16.704 &       0.021 \\
2459115.44 & 2020-09-22 & 16.32 &      \(r^{\prime}\) &          16.833 &       0.025 \\
2459121.45 & 2020-09-28 & 22.23 &      \(r^{\prime}\) &          17.228 &       0.040 \\
2459122.40 & 2020-09-29 & 23.16 &      \(r^{\prime}\) &          17.296 &       0.025 \\
2459127.39 & 2020-10-04 & 28.07 &      \(r^{\prime}\) &          17.566 &       0.024 \\
2459133.42 & 2020-10-10 & 34.00 &      \(r^{\prime}\) &          17.707 &       0.038 \\
2459139.42 & 2020-10-16 & 39.90 &      \(r^{\prime}\) &          17.916 &       0.014 \\
2459143.38 & 2020-10-20 & 43.80 &      \(r^{\prime}\) &          18.030 &       0.019 \\
2459144.37 & 2020-10-21 & 44.77 &      \(r^{\prime}\) &          18.037 &       0.182 \\
2459145.33 & 2020-10-22 & 45.71 &      \(r^{\prime}\) &          18.092 &       0.030 \\
2459147.33 & 2020-10-24 & 47.67 &      \(r^{\prime}\) &          18.235 &       0.033 \\
2459149.37 & 2020-10-26 & 49.68 &      \(r^{\prime}\) &          18.123 &       0.173 \\
2459150.36 & 2020-10-27 & 50.66 &      \(r^{\prime}\) &          18.075 &       0.012 \\
2459162.36 & 2020-11-08 & 62.46 &      \(r^{\prime}\) &          18.531 &       0.023 \\
2459174.31 & 2020-11-20 & 74.21 &      \(r^{\prime}\) &          18.744 &       0.036 \\
2459175.34 & 2020-11-21 & 75.22 &      \(r^{\prime}\) &          18.752 &       0.040 \\
2459177.33 & 2020-11-23 & 77.17 &      \(r^{\prime}\) &          18.850 &       0.045 \\
\hline
\multicolumn{3}{l}{$^*$\footnotesize{Time since $B$-band maximum (2459098.84)}.}
\end{longtable}
}
\afterpage{
\begin{longtable}[hbt!]{ccccccc}
\caption{$gri$ magnitudes of SN~2020sck from DOT.} \label{tab:dot_photlog} \\

\hline \multicolumn{1}{c}{JD} & \multicolumn{1}{c}{Date} & \multicolumn{1}{c}{Phase$^*$} & \multicolumn{1}{c}{Filter} & \multicolumn{1}{c}{Magnitude (mag)} & \multicolumn{1}{c}{Error (mag)} \\ \hline 
\endfirsthead

\multicolumn{6}{c}%
{{\bfseries \tablename\ \thetable{} -- continued from previous page}} \\
\hline \multicolumn{1}{c}{JD} & \multicolumn{1}{c}{Date} & \multicolumn{1}{c}{Phase$^*$} & \multicolumn{1}{c}{Filter} & \multicolumn{1}{c}{Magnitude (mag)} & \multicolumn{1}{c}{Error (mag)} \\ \hline 
\endhead

\hline \multicolumn{6}{|r|}{{Continued on next page}} \\ \hline
\endfoot

\hline \hline
\endlastfoot
2459226.40 & 2021-01-11 & 127.56 &     \(g\)   & 19.75  & 0.05 \\
2459226.41 & 2021-01-11 & 127.55 &     \(r\)   & 18.84  & 0.12  \\
2459226.40 & 2021-01-11 & 127.56 &     \(i\)   & 18.98  & 0.13 \\
\hline
\multicolumn{3}{l}{$^*$\footnotesize{Time since $B$-band maximum (2459098.84)}.}
\end{longtable}
}
\begin{table*}
\centering
\setlength{\tabcolsep}{3pt}
\caption{Log of spectroscopic observations of SN~2020sck from HCT.}
\label{tab:speclog}
\begin{tabular}{c c c c}
\hline \hline
    JD &     Date &  Phase* & Range  \\
(2459000+) &   &  (d)    & (\AA)         \\
 \hline \noalign{\smallskip}    
 93.31 &  2020-08-31  &  -5.53 & 3500-7800; 5200-9100   \\
 94.27 &  2020-09-01  &  -4.57 & 3500-7800; 5200-9100 \\
 95.34 &  2020-09-02  &  -3.50 & 3500-7800; 5200-9100 \\
100.24 &  2020-09-07  &   1.40 & 3500-7800; 5200-9100 \\
101.45 &  2020-09-08  &   2.61 & 3500-7800 \\
104.35 &  2020-09-11  &   5.51 & 3500-7800; 5200-9100 \\
106.39 &  2020-09-13  &   7.55 & 3500-7800; 5200-9100 \\
109.28 &  2020-09-16  &  10.44 & 3500-7800; 5200-9100 \\
113.18 &  2020-09-20  &  14.34 & 3500-7800 \\
117.31 &  2020-09-24  &  18.47 & 3500-7800; 5200-9100 \\
120.40 &  2020-09-27  &  21.56 & 3500-7800 \\
122.39 &  2020-09-29  &  23.55 & 3500-7800; 5200-9100 \\
126.34 &  2020-10-03  &  27.50 & 3500-7800; 5200-9100 \\
\noalign{\smallskip} \hline
\multicolumn{3}{l}{$^*$\footnotesize{Time since $B$-band maximum (JD 2459098.84)}.}
\end{tabular}
\end{table*}

\end{document}